\documentclass[12pt,a4paper,superscriptaddress,preprint,longbibliography]{revtex4-1}
\usepackage[dvips]{graphicx} 

\usepackage{amssymb,amsmath}
\usepackage{bm}
\usepackage{color}

\usepackage{supertabular}
\usepackage{multirow}
\usepackage{bigstrut}

\usepackage{dcolumn}
\newcolumntype{Y}{>{\raggedleft\arraybackslash}p{15mm}}
\newcolumntype{Z}{>{\raggedleft\arraybackslash}p{30mm}}

\usepackage{ulem}

\begin{document}

\title{Configuration model for correlation matrices preserving the node strength}
\date{\today}
\author{Naoki Masuda}
\affiliation{Department of Engineering Mathematics,
Merchant Venturers Building, University of Bristol,
Woodland Road, Clifton, Bristol BS8 1UB, United Kingdom}
\email{naoki.masuda@bristol.ac.uk}
\author{Sadamori Kojaku}
\affiliation{Department of Engineering Mathematics,
Merchant Venturers Building, University of Bristol,
Woodland Road, Clifton, Bristol BS8 1UB, United Kingdom}
\affiliation{CREST, JST, Kawaguchi Center Building, 4-1-8,
Honcho, Kawaguchi-shi, Saitama 332-0012, Japan}
\author{Yukie Sano}
\affiliation{Faculty of Engineering, Information and Systems,
University of Tsukuba, 1-1-1 Tennodai, Tsukuba, Ibaraki, 305-8577 Japan}

\begin{abstract}
Correlation matrices are a major type of multivariate data. To examine properties of a given correlation matrix, a common practice is to compare the same quantity between the original correlation matrix and reference correlation matrices, such as those derived from random matrix theory, that partially preserve properties of the original matrix. We propose a model to generate such reference correlation and covariance matrices for the given matrix. Correlation matrices are often analysed as networks, which are heterogeneous across nodes in terms of the total connectivity to other nodes for each node. Given this background, the present algorithm generates random networks that preserve the expectation of total connectivity of each node to other nodes, akin to configuration models for conventional networks. Our algorithm is derived from the maximum entropy principle. We will apply the proposed algorithm to measurement of clustering coefficients and community detection, both of which require a null model to assess the statistical significance of the obtained results.
\end{abstract}

\maketitle

\section{Introduction\label{sec:introduction}}

Correlation matrices are a major form of multivariate data in various domains. Examples include financial time series \cite{Laloux1999PhysRevLett,Plerou1999PhysRevLett}, behavioural and questionnaire data in psychology \cite{Borsboom2013AnnuRevClinPsychol}, genetic interactions \cite{Horvath2008PlosComputBiol,Junker2008book,Vidal2011Cell}, neuroscience \cite{Chapin2004NatNeurosci,Schneidman2006Nature,Bullmore2009NatRevNeurosci} and climate science \cite{Tsonis2006BullAmerMeteorolSoc}. Although pairwise correlation does not always reflect physical connection or direct interaction between two entities, correlation matrices, whose entries represent the strength of correlation between a pair of entities (which we call nodes in the rest of the paper), are conventionally used as a relatively inexpensive substitute for direct connection. 

Major analysis tools for correlation matrix data include principal component analysis \cite{Jolliffe2002book}, factor analysis \cite{Harman1976book}, Markowitz's portfolio theory in mathematical finance \cite{Markowitz1952JFinance} and random matrix theory \cite{Laloux1999PhysRevLett,Plerou1999PhysRevLett}.
A more recent approach to correlational data is network analysis. With this approach, the first task is usually to either threshold on the value of the pairwise correlation to define an unweighted (i.e., binary) network or adopt the value of the pairwise correlation as the edge weight to define a weighted network. Then, one examines properties of the obtained network. Network analysis provides information that is different from the information obtained with other methods, such as the distance between two nodes, centrality (i.e., importance) of the nodes according to various criteria and network motifs (i.e., overrepresented small subgraphs) 
\cite{Wasserman1994book,Newman2010book,Barabasi2016book}. Network analysis of correlation matrices is common across disciplines \cite{Bullmore2009NatRevNeurosci,Tsonis2006BullAmerMeteorolSoc,Horvath2008PlosComputBiol,Junker2008book,Vidal2011Cell,Borsboom2013AnnuRevClinPsychol,Eguiluz2005PhysRevLett,WangWang2009HumanBrainMapping,KimLee2002JPhysSocJapan,Bonanno2003PhysRevE,Namaki2011PhysicaA,Jordan2004MolBiolEvol,Bredel2005CancerRes}. 

However, there are fundamental technical problems in applying standard methods of network analysis to correlation matrix data. First, correlation networks tend to suffer from type 1 errors (i.e., false positives) because pairwise correlation does not differentiate between direct effects (i.e., nodes $v_1$ and $v_2$ are correlated because they directly interact) and indirect effects (i.e., $v_1$ and $v_2$ are correlated because nodes $v_1$ and $v_3$ interact and $v_2$ and $v_3$ interact) \cite{Barzel2013NatBiotechnol,Fiecas2013Neuroimage,Hinne2015PlosComputBiol}. Second, when analysing a correlation matrix as an unweighted network, no consensus exists regarding the choice of the threshold value despite the evidence that results are sensitive to the threshold (e.g., Ref.~\cite{Devicofallani2017PlosComputBiol}).
Third, whereas thresholding is claimed to mitigate uncertainty on weak links and enhance interpretability of network-analysis results \cite{Vidal2011Cell,Devicofallani2017PlosComputBiol}, thresholding discards potentially important information contained in the values of the correlation coefficient \cite{Rubinov2011Neuroimage}.
Last, even if we do not carry out thresholding and treat a correlation matrix as a weighted network, the problem of type-1 errors remains and it is unclear how to deal with negatively weighted edges.

We consider that these shortcomings inherent in correlation network analysis owe to the paucity of network-analysis tools tailored to correlation matrices. Not just being symmetric, correlation matrices are a special type of matrices in that they are positive semidefinite (i.e., all eigenvalues are nonnegative), they are dense and the range of the entries is confined between $-1$ and $1$ \cite{Holmes1991SiamJMatAnalAppl}. Furthermore, the node $i$'s weighted degree in a correlation matrix represents the correlation between node $i$ and the average of the signal over all nodes, which is somewhat non-intuitive and affects analysis of correlation networks \cite{Macmahon2015PhysRevX}. We do have algorithms to partition correlation networks into communities \cite{Macmahon2015PhysRevX}, calculate their clustering coefficients
\cite{Masuda2017submitted} and detect change points in time-varying correlation networks
 \cite{Barnett2016SciRep}. These algorithms are tailored to correlation matrix input. However, these analysis tools were proposed only recently, and analysis tools for correlation matrices as networks still seem to be in their infancy.

In the present paper, we propose a configuration model for correlation matrices and showcase its use as the null model in measuring the clustering coefficient and community structure. In general, a null model of networks generates randomised networks that preserve some but not all features of the given network. Then, one compares a property in question calculated for the given network and that calculated for sample networks generated by the null model to assess whether the property of the given network is significant relative to that of the null model \cite{Porter2009AmerMathSoc,Fosdick2017SiamRev}.
Null models available for correlation matrices include the identity matrix \cite{Macmahon2015PhysRevX}, 
Laguerre ensembles \cite{Laloux1999PhysRevLett} or Gaussian orthogonal ensemble \cite{Plerou1999PhysRevLett} of random matrix theory,  
Hirschberger-Qu-Steuer (H-Q-S) algorithm \cite{Hirschberger2007EurJOperRes} and correlation matrices reconstructed from noise eigenmodes corresponding to small eigenvalues of the correlation matrix (and the largest eigenmode in a different variant) \cite{Macmahon2015PhysRevX}.

\begin{figure}
\begin{center}
% Edits by Kojaku
\includegraphics[width=\hsize]{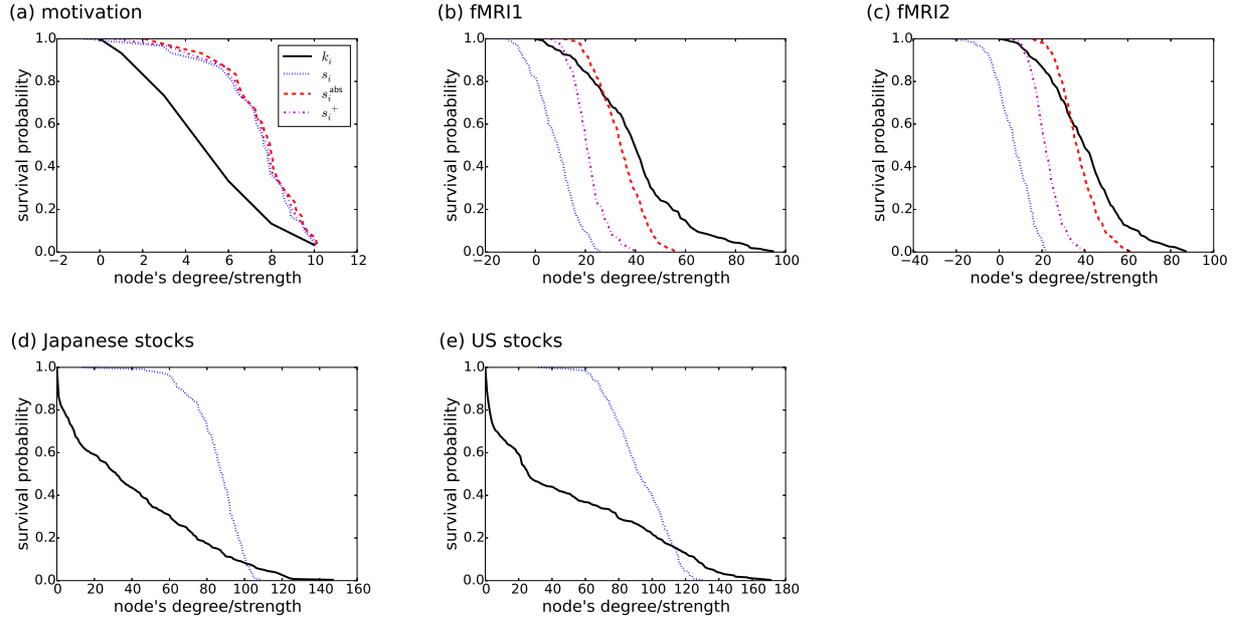} 
\caption{Distributions of the degree ($k$; solid lines) and the three types of node strength.
The strength of node $i$ is defined as (i) $s_i = \sum_{j=1; j\neq i}^N \rho_{ij}$
(dotted lines), (ii) $s_i^{\rm abs} = \sum_{j=1; j\neq i}^N \left| \rho_{ij}\right|$ (dashed lines) and (iii) $s_i^+ = \sum_{j=1; j\neq i; \rho_{ij}>0}^N \rho_{ij}$ (dot-dashed lines). In (d) and (e), it holds that $s_i \approx s_i^{\rm abs} \approx s_i^+$ ($1\le i\le N$) because $\rho^{\rm org}_{ij}\ge 0$ for all but three pairs of nodes in (d) and all but one pair of nodes in (e). Therefore, we did not plot $s_i^{\rm abs}$ or $s_i^+$.
(a) Motivation. (b) fMRI1. (c) fMRI2. (d) Japanese stocks. (e) US stocks.}
\label{fig:distribution of k and s}
\end{center}
\end{figure}

For conventional networks (i.e., networks not derived from correlation matrices), 
heterogeneity in the degree distribution is a common feature in empirical data \cite{Newman2010book,Barabasi2016book}.
Configuration models are probably the most often used class of null models and generate random networks under the constraint that generated networks conserve the (expected) degree of each node in the original network \cite{Molloy1995RandomStructAlgor,Goh2001PhysRevLett,Chung2002PNAS,Fosdick2017SiamRev}.
Heterogeneous degree distributions have also been observed for
correlation networks of the brain \cite{Eguiluz2005PhysRevLett,WangWang2009HumanBrainMapping}, financial data \cite{KimLee2002JPhysSocJapan,Namaki2011PhysicaA} and gene coexpression \cite{Jordan2004MolBiolEvol,Bredel2005CancerRes} (also see Fig.~\ref{fig:distribution of k and s}). However, none of the aforementioned null models for correlation matrices is intended to preserve the degree or strength (i.e., weighted degree) of the nodes in correlation matrices, which motivates the present paper.

% Other studies on canonical ensembles of networks (not mentioning entropy maximisation): \cite{ParkNewman2003PhysRevE,Garlaschelli2006PhysRevE}

\section{Maximum entropy model for correlation matrices preserving the expected strength of each node\label{sec:model}}

We propose a configuration model for covariance matrices.
We work with covariance matrices rather than correlation matrices due to analytical tractability of the former.
However, in practice, we usually analyse correlation matrices rather than covariance matrices because the latter is un-normalised. Therefore, we need a configuration model for correlation matrices. Given this situation, we will explain applications of our algorithm to correlation matrices in section~\ref{sub:to cor} first. This discussion gives two conditions that constrain the configuration model for covariance matrices, which will be developed in section~\ref{sub:formulation and algorithm}. MATLAB
% and python
codes for estimating the configuration model are available at Github \cite{code}.%(\url{https://github.com/naokimas/config_corr}).

\subsection{Conservation of the node strength in correlation matrices\label{sub:to cor}}

Denote by $\Sigma^{\rm org}$ the $N\times N$ covariance matrix given as input
and by $\Sigma^{\rm con}$ the $N\times N$ covariance matrix obtained from the configuration model. We will explain how to calculate $\Sigma^{\rm con}$ from 
$\Sigma^{\rm org}$ in section~\ref{sub:formulation and algorithm}. When the input is a correlation matrix, denoted by $\rho^{\rm org}$, our
aim is to ensure that the expected strength of each node of the correlation matrix generated by the configuration model, denoted by $\rho^{\rm con}$, is similar to that of $\rho^{\rm org}$.

To this end, we start by discussing the relationship between the entries of the covariance matrix and the node strength in the corresponding correlation matrix.
The Pearson correlation between nodes $i$ and $j$ is given by
\begin{equation}
\rho_{ij} = \frac{\Sigma_{ij}}{\sqrt{\Sigma_{ii} \Sigma_{jj}}},
\label{eq:rho}
\end{equation}
where $\rho = (\rho_{ij})$ is the correlation matrix corresponding to a covariance matrix $\Sigma$.
A direct equivalent of the strength of node $i$ in the correlation matrix, denoted by $s_i$, is given by
\begin{equation}
s_i \equiv \sum_{j=1; j\neq i}^N\rho_{ij} = \frac{1}{\sqrt{\Sigma_{ii}}}
\sum_{j=1; j\neq i}^N \frac{\Sigma_{ij}}{\sqrt{\Sigma_{jj}}}.
\label{eq:s_i}
\end{equation}
Equation~\eqref{eq:s_i} indicates that, if each diagonal element of $\Sigma^{\rm con}$ and the row sum of the off-diagonal elements of $\Sigma^{\rm con}$ for each row are equal to those for $\Sigma^{\rm org}$, the configuration model, which will be formulated in section~\ref{sub:formulation and algorithm}, roughly conserves $s_i$ ($1\le i\le N$) of the input correlation matrix. Therefore, in our configuration model, we will impose that the expectation of
$\Sigma^{\rm con}_{ii}$ and $\sum_{j=1; j\neq i}^N \Sigma^{\rm con}_{ij}$ are equal to $\Sigma^{\rm org}_{ii}$ and $\sum_{j=1; j\neq i}^N \Sigma^{\rm org}_{ij}$, respectively, for each $i$ ($1\le i\le N$).

In fact, Eq.~\eqref{eq:s_i} implies that, even under these constraints, the expected node strength for $\rho^{\rm con}$ is not generally equal to the node strength for $\rho^{\rm org}$. The discrepancy would be large if the autocovariance, $\Sigma^{\rm org}_{jj}$, which appears in the denominator in Eq.~\eqref{eq:s_i}, heavily depends on $j$. In contrast, if
$\Sigma^{\rm org}_{jj}$ is independent of $j$, then 
$\sum_{j=1; j\neq i}^N \Sigma^{\rm con}_{ij} = \sum_{j=1; j\neq i}^N \Sigma^{\rm org}_{ij}$ ($1\le i\le N$) guarantees that the configuration model conserves $s_i$ of the correlation matrix for each $i$. A correlation matrix is a covariance matrix (therefore allowed as input to our algorithm developed in section~\ref{sub:formulation and algorithm}) and its diagonal elements are independent of the node (i.e., equal to 1 for each node). Therefore, when a correlation matrix is fed to our algorithm, we expect that the output conserves the node strength to a high accuracy.

\begin{figure}
\begin{center}
\includegraphics[width=16cm]{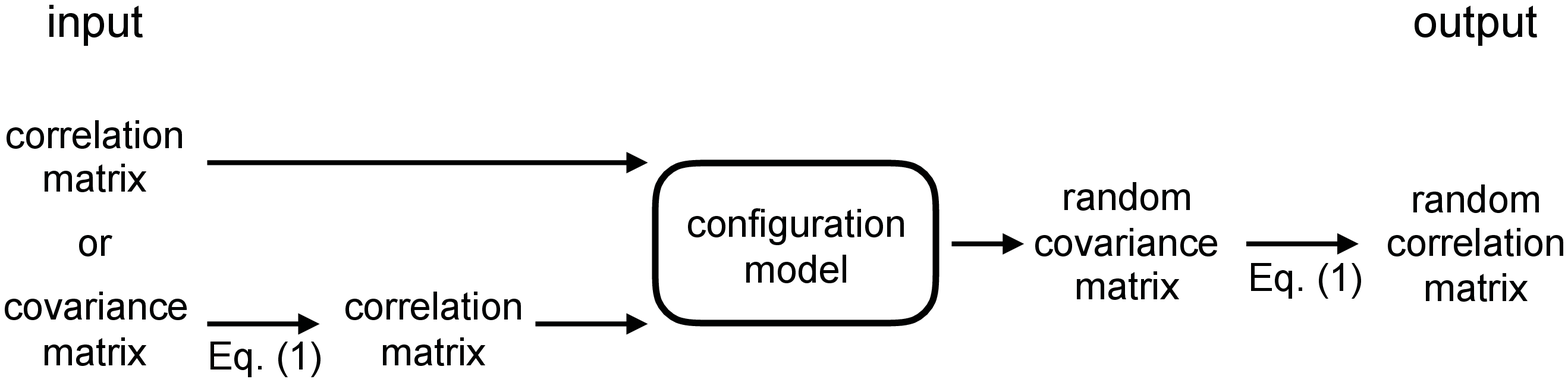}
\caption{Flow of the algorithm for generating random correlation matrices. If the input is a covariance matrix, we first transform it to the correlation matrix and feed it to the configuration model. If the input is a correlation matrix, we directly feed it to the configuration model. Because the output of the configuration model is a random covariance matrix (or samples generated from it), we transform it to the correlation matrix, which is the final output.}
\label{fig:schem}
\end{center}
\end{figure}

The flow of the algorithm is shown in Fig.~\ref{fig:schem}. If the original data are a covariance matrix, we first transform it to a correlation matrix, $\rho^{\rm org}$, using Eq.~\eqref{eq:rho}.
Then, we submit $\rho^{\rm org}$, which is a covariance matrix, to our algorithm. Because the input covariance matrix (i.e., $\rho^{\rm org}$) has uniform diagonal elements (i.e., all equal to 1), we expect that the algorithm approximately conserves the node's strength of the correlation matrix. 
The output of the algorithm is a covariance matrix whose expectation of each diagonal element is equal to 1. Note that each diagonal element of the output covariance matrix is not generally equal to 1. Finally, we transform the output covariance matrix to the correlation matrix, which is denoted by $\rho^{\rm con}$, using Eq.~\eqref{eq:rho}.

\subsection{Maximum entropy formalism and the gradient descent algorithm\label{sub:formulation and algorithm}}

Assume a covariance matrix $\Sigma^{\rm org}$ as input.
We generate random covariance matrices that conserve the expectation of the row sum of the off-diagonal elements of $\Sigma^{\rm org}$ in each row and the expectation of each diagonal element, i.e., the auto-covariance of each node, of $\Sigma^{\rm org}$. We achieve this goal by standing on the maximum entropy principle, with which one generates the distribution with the largest entropy under certain constraints
\cite{Jaynes1957PhysRev}. For conventional networks, the maximum entropy principle has been used for generating unweighted \cite{ParkNewman2004PhysRevE,Squartini2011NewJPhys,Almog2015NewJPhys,Squartini2015NewJPhys,Squartini2017book} and weighted 
\cite{Garlaschelli2009PhysRevLett,Squartini2011NewJPhys,Mastrandrea2014NewJPhys,Almog2015NewJPhys,Cimini2015PhysRevE,Squartini2015NewJPhys,Squartini2017book}
networks (also see \cite{Bianconi2008EPL,Anand2009PhysRevE,Bianconi2009PhysRevE}).
However, networks generated by these algorithms are not correlation or covariance matrices in general.

Denote by $N$ the number of elements, which we refer to as nodes according to the terminology of networks. We generate $N \times N$ covariance matrices of the following form:
\begin{equation}
\Sigma^{\rm con} = (\Sigma^{\rm con}_{ij}) = \frac{1}{L} X X^{\top},
\label{eq:C with X}
\end{equation}
where $X= (x_{ij})$ is an $N\times L$ real matrix and $\top$ denotes the transposition. Because a covariance matrix is positive semidefinite, its eigendecomposition implies that any given covariance matrix can be written in the form of Eq.~\eqref{eq:C with X} when $L$ is larger than or equal to the number of positive eigenvalues of $\Sigma^{\rm con}$.
%
% \cite{Davies2000BitNumerMath}.
%
Because 
\begin{equation}
\Sigma^{\rm con}_{ij} = \frac{1}{L} \sum_{\ell=1}^L x_{i\ell} x_{j\ell},
\end{equation}
matrix $\Sigma^{\rm con}$ is interpreted as the sample covariance matrix when the $i$th data vector (e.g., time series in discrete time or a feature vector) is given by $\{x_{i1}, \ldots, x_{iL}\}$.

We will determine a distribution of matrix $X$, which we denote by $p(X)$. Under the maximum entropy principle, we maximise
\begin{align}
H(X) \equiv& - \int p(X) \ln p(X) {\rm d}X
+ \sum_{i=1}^N \alpha_i \left[ \int \Sigma^{\rm con}_{ii} p(X) {\rm d}X - \Sigma^{\rm org}_{ii} \right]\notag\\
&+ \sum_{i=1}^N \beta_i \left[ \int \sum_{j=1; j\neq i}^N \Sigma^{\rm con}_{ij} p(X) {\rm d}X - \sum_{j=1; j\neq i}^N \Sigma^{\rm org}_{ij} \right],
\label{eq:entropy with Lagrange multipliers}
\end{align}
where $\alpha_i$ and $\beta_i$ are Lagrange multipliers.
By taking the functional derivative of Eq.~\eqref{eq:entropy with Lagrange multipliers} with respect to $p(X)$ and setting it to zero, we obtain
\begin{align}
p(X) \propto&
\exp\left[ \frac{1}{L} \sum_{\ell=1}^L \left( \sum_{i=1}^N \alpha_i x_{i\ell}^2 + \sum_{i=1}^N \beta_i \sum_{j=1; j\neq i}^N x_{i\ell} x_{j\ell}\right)\right]\notag\\
=& 
\prod_{\ell=1}^L \exp\left[ \frac{1}{L} \left(\sum_{i=1}^N \alpha_i x_{i\ell}^2 + \sum_{i=1}^N \beta_i \sum_{j=1; j\neq i}^N x_{i\ell} x_{j\ell}\right) \right]\notag\\
=& \prod_{\ell=1}^L \exp\left[-\frac{1}{2}
\bm x_{\ell}^{\top} \Sigma^{-1} \bm x_{\ell} \right],
\label{eq:p(X) max ent def}
\end{align}
where $\bm x_{\ell} = (x_{1\ell}, \ldots, x_{N,\ell})^{\top}$ and
\begin{equation}
\Sigma^{-1} = - \frac{1}{L}\begin{pmatrix}
2\alpha_1 & \beta_1+\beta_2 & \beta_1+\beta_3 & \cdots & \beta_1+\beta_{N-1} & \beta_1+\beta_N\\
\beta_2+\beta_1 & 2\alpha_2 & \beta_2+\beta_3 & \cdots & \beta_2+\beta_{N-1} & \beta_2+\beta_N\\
& & \vdots & & &\\
\beta_N+\beta_1 & \beta_N+\beta_2 & \cdots & \cdots & \beta_N+\beta_{N-1} & 2\alpha_N
\end{pmatrix}.
\label{eq:precision matrix}
\end{equation}
Therefore, $p(X)$ is given by a multivariate normal distribution, i.e.,
\begin{equation}
p(X) = \prod_{\ell=1}^L \frac{1}{\sqrt{(2\pi)^N |\Sigma |}} \exp\left[-\frac{1}{2}
\bm x_{\ell}^{\top} \Sigma^{-1} \bm x_{\ell} \right],
\label{eq:p(X) multivariate normal}
\end{equation}
with which one draws
$\bm x_{\ell}$ for each $\ell$ from the $N$-variate multivariate normal distribution with mean zero and precision matrix $\Sigma^{-1}$, independently for each $\ell$. Note that $\Sigma$ is the covariance matrix for the estimated multivariate normal distribution.

To numerically determine the precision matrix, we reparametrise Eq.~\eqref{eq:precision matrix} as
\begin{equation}
\Sigma^{-1} = \begin{pmatrix}
\alpha_1+2\beta_1 & \beta_1+\beta_2 & \beta_1+\beta_3 & \cdots & \beta_1+\beta_{N-1} & \beta_1+\beta_N\\
\beta_2+\beta_1 & \alpha_2+2\beta_2 & \beta_2+\beta_3 & \cdots & \beta_2+\beta_{N-1} & \beta_2+\beta_N\\
& & \vdots & & &\\
\beta_N+\beta_1 & \beta_N+\beta_2 & \cdots & \cdots & \beta_N+\beta_{N-1} & \alpha_N+2\beta_N
\end{pmatrix}
\label{eq:precision matrix 2}
\end{equation}
without loss of generality. 
We infer $\Sigma^{-1}$ by running the following gradient descent algorithm.

Equation~\eqref{eq:p(X) multivariate normal} leads to
\begin{equation}
\frac{\partial}{\partial \Sigma^{-1}_{ij}}\log p(X) =
\frac{L}{2} \left( \Sigma_{ij} - \Sigma_{ij}^{\rm org}\right).
\end{equation}
Therefore, the gradient descent learning rule for $\alpha_i$ and $\beta_i$ to maximise $H(X)$ is given by
\begin{align}
\alpha_i^{\rm new} =& \alpha_i^{\rm old} + \epsilon \left(\Sigma_{ii} - \Sigma^{\rm org}_{ii} \right),
\label{eq:gradient descent alpha_i}\\
\beta_i^{\rm new} =& \beta_i^{\rm old} + 2 \epsilon \left(\sum_{j=1}^N \Sigma_{ij} - \sum_{j=1}^N \Sigma^{\rm org}_{ij} \right),
\label{eq:gradient descent beta_i}
\end{align}
where $1\le i\le N$ and $\epsilon$ is the learning rate.
We refer to Eq.~\eqref{eq:p(X) multivariate normal} with the optimised $\alpha_i$ and $\beta_i$ values as the configuration model for correlation matrices.
In the numerical simulations in Section~\ref{sec:numerical results}, we set
$\epsilon = 10^{-4}$. We remark that the gradient descent algorithm, and hence the obtained precision matrix, does not depend on our choice of $L$.

\subsection{Choice of $L$\label{sub:choice of L}}

A covariance matrix $\Sigma^{\rm con}$ obtained from our configuration model obeys a Wishart distribution with degree of freedom $L$, denoted by $W_N(L, \Sigma)$. The mean of each element of $\Sigma^{\rm con}$ is given by $\Sigma$ and the variance of $\Sigma^{\rm con}_{ij}$ ($1\le i, j\le N$) is given by
$(\Sigma_{ij}^2 + \Sigma_{ii}\Sigma_{jj})/L$ \cite{GuptaNagar2000book,Kollo2005book}.
Therefore, $L$ controls the amount of fluctuations in covariance matrices generated by the algorithm. In the limit of $L\to\infty$, the configuration model always produces covariance matrix $\Sigma$, in which the strength of each node and each diagonal element agree with those of the input covariance matrix, $\Sigma^{\rm org}$. If $L$ is finite, the configuration model produces covariance matrices that differ from sample to sample.

We set $L$ to the length of the original data based on which the covariance or correlation matrix is calculated (e.g., the length of time series, number of participants in an experiment, or dimension of the feature vector). If the length of the original data is unknown, we propose to set $L$ to the number of positive eigenvalues of
$\Sigma^{\rm org}$ because it is the smallest value of $L$ with which the configuration model may preserve the rank of the input covariance matrix in addition to the node strength. We remark that our gradient descent algorithm often fails when 
$\Sigma^{\rm org}$ is not of full rank (hence $L<N$). In the following sections, we use empirical data whose $L$ value is known and $L > N$.

\subsection{Uniformity of samples\label{sub:non-uniform}} 

By maximising the entropy in terms of $p(X)$, our configuration model does not maximise the entropy in terms of the distribution of all possible positive semidefinite matrices, which qualify as covariance matrices. Therefore, our model is biased in the space of all possible positive semidefinite matrices. However, we consider that it is rather realistic to formulate the maximum entropy principle in terms of $p(X)$ because empirical covariance matrices are usually calculated from Eq.~\eqref{eq:C with X}, where $X$ is raw data.

\section{Numerical Results\label{sec:numerical results}}

\subsection{Data}

We use five empirical correlation matrices to compare different methods. In all cases,
the empirical correlation matrix, $\rho^{\rm org}$, is calculated as the Pearson correlation coefficient between pairs of multidimensional measurements or time series.

The first correlation matrix is based on psychological questionnaires with $N=30$ question items. We refer to this data set as the motivation data.
The questionnaires consist of three scales (i.e., inventories) of academic motivation at school. The first scale is the so-called Achievement Goal Questionnaire (18 items) \cite{Elliot1997JPersSocPsychol}, which assesses students' mastery goals (i.e. goals to master a task), performance-approach goals (i.e. goals
to outperform others) and performance-avoidance goals (i.e. goals not
to be outperformed by others) in a class. The second scale is a
shortened version (six items) of an intrinsic motivation scale used in Ref.~\cite{Elliot1997JPersSocPsychol}, which assesses students' intrinsic motivation or
enjoyment in a class. The last one is an academic self-concept scale
(six items) \cite{Ichihara2004WCBC}, which assesses students'
competence belief about a class. School children responded to these
questionnaire items on a five-point Likert scale (1, strongly disagree --
5, strongly agree).
The Pearson correlation coefficient between each pair of items is calculated from responses from $L=686$ persons. The correlation matrix is available as Supplementary Material.

Two correlation matrices are obtained from multivariate time series of functional magnetic resonance imaging (fMRI) signals in the brain. Each correlation matrix is derived from a human participant.
The data are collected from the Human Connectome Project \cite{Vanessen2012Neuroimage}. For each of the two participants, we extract time series at $N=264$ locations whose coordinates are determined in a previous study \cite{Power2011Neuron}. The pairwise correlation is calculated based on fMRI time series of length $L=4,760$. We refer to the data from the two participants as fMRI1 and fMRI2. Details of the preprocessing procedures are explained in Appendix~\ref{app:preprocessing}.

We also use two correlation matrices obtained from time series of
the logarithmic return of the daily closing prices in the Japanese and US stock markets.
For the Japanese data, we use the $N = 264$ stocks belonging to the first section of the Tokyo Stock Exchange provided by Nikkei NEEDS \cite{Nikkei2017Needs}. 
We limit ourselves to the stocks that have transactions on every trading day between 12 March 1996 and 29 February 2016, yielding $4,904$ trading days in total. 
For the US data, we obtain the $N = 325$ stocks from the list of the Standard \& Poor's 500 index using Mathematica's FinancialData package \cite{Mathematica2014}. 
We limit ourselves to the stocks that have transactions on every trading day between 3 January 1996 and 24 February 2017, yielding $5,324$ trading days in total. For each stock, we convert the time series of the stock price into that of the logarithmic return by $x_{it}^{\rm org}
 = \log (y_{i,t+1}^{\rm org} / y_{i,t}^{\rm org})$, where $y_{i,t}^{\rm org}$ is the closing price of the $i$th stock on the $t$th day, and $x_{it}^{\rm org}$ is the corresponding logarithmic return. The length of $\{y_{i,t}\}$ is equal to $L=4,903$ and $L=5,323$ for the Japanese and US data sets, respectively.

\subsection{Degree and strength distributions for the empirical correlation matrices and networks}

The motivation behind our configuration model is that the node strength value depends on nodes. Otherwise, the previously proposed models to generate random correlation or covariance matrices \cite{Laloux1999PhysRevLett,Plerou1999PhysRevLett,Hirschberger2007EurJOperRes,Macmahon2015PhysRevX} would probably suffice.
Therefore, in this section we measure the distribution of the node's degree and strength in the empirical networks. To calculate the degree of each node $i$, which is denoted by $k_i$, we binarise the correlation matrix to create an unweighted network. For this purpose, we threshold on the pairwise correlation value to make the edge density equal to 0.15, which is an arbitrary choice. To calculate the strength of each node $i$, we consider weighted networks obtained without the thresholding on the pairwise correlation value.
For the weighted networks, we define the node strength by either (i) the sum of the off-diagonal elements of the correlation matrix, denoted by $s_i$; (ii) the same sum but using the absolute value of the correlation, denoted by $s_i^{\rm abs}$; or (iii) the same sum but discarding negative correlation values, denoted by $s_i^+$.

The survival probability (i.e., probability that a quantity is larger than or equal to the specified value) of the degree and the three types of node strength are shown in 
Fig.~\ref{fig:distribution of k and s} for each empirical network.
As briefly mentioned in Section~\ref{sec:introduction}, the degree and strength are to some extent heterogeneous across nodes, although the distributions are not long-tailed.

\subsection{Distribution of eigenvalues}

Random matrix theory is a useful tool to formulate null models of correlation matrices
\cite{Laloux1999PhysRevLett,Plerou1999PhysRevLett}. MacMahon and Garlaschelli proposed a null model of a correlation matrix, which we denote by $\langle \rho^{\rm MG2} \rangle$ \cite{Macmahon2015PhysRevX}. Matrix $\langle \rho^{\rm MG2} \rangle$ preserves the eigenmodes of the input correlation matrix, $\rho^{\rm org}$, that correspond to small eigenvalues, i.e., those contained in the spectrum of a correlation matrix constructed from $N$ completely random time series of length $L$.
Their other null model, which we denote by $\langle \rho^{\rm MG3}\rangle$, preserves
the eigenmode corresponding to the largest eigenvalue of $\rho^{\rm org}$ in addition to the noisy eigenmodes used in $\langle \rho^{\rm MG2}\rangle$. See Appendix~\ref{app:MG} for the definition of $\langle \rho^{\rm MG2} \rangle$ and $\langle \rho^{\rm MG3} \rangle$. To relate the present configuration model to random matrix theory, we investigate the eigenvalue distribution for our configuration model in this section.

\begin{figure}
\begin{center}
% Edits by Kojaku
\includegraphics[width=\hsize]{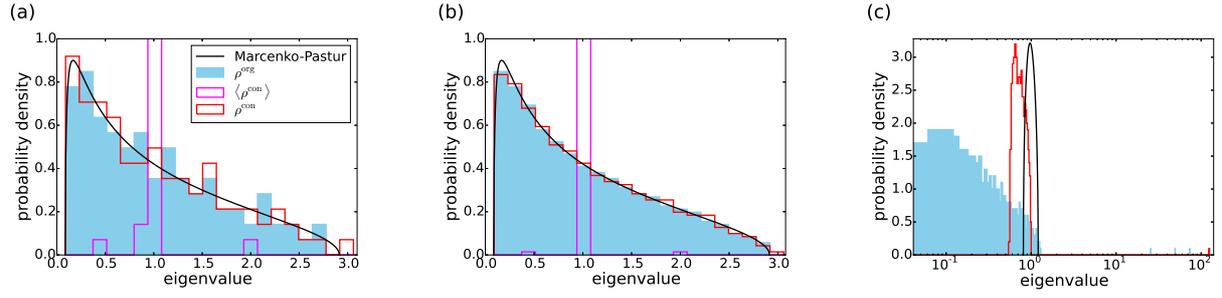}
\caption{Density of the eigenvalues of a random correlation matrix, denoted by $\rho^{\rm org}$, and the corresponding configuration model, denoted by $\rho^{\rm con}$. The black solid lines represent the Marcenko-Pastur distribution. (a) $N=100$ and $L=200$ without community structure. (b) $N=500$ and $L=1,000$ without community structure. (c) $N=500$ and $L=1,000$ with community structure.}
\label{fig:eigs random}
\end{center}
\end{figure}

We first generate a correlation matrix, $\rho^{\rm org}$, from $N$ completely independent normally distributed time series of length $L$. With this random correlation matrix as input, we estimate the configuration model. Then, we generate a sample correlation matrix, denoted by $\rho^{\rm con}$, from the estimated configuration model.
With $N=100$ and $L=200$, the distribution of the eigenvalues of the original correlation matrix and that of a sample correlation matrix generated by the configuration model are shown in skyblue and red in Fig.~\ref{fig:eigs random}(a), respectively. The figure suggests that the two distributions are similar. Furthermore, both distributions are similar to the theoretical distribution for the completely random correlation matrix called the Marcenko-Pastur (also called Sengupta-Mitra) distribution given by
\begin{equation}
p(\lambda) = \begin{cases}
\frac{L}{N} \frac{\sqrt{(\lambda_+-\lambda)(\lambda-\lambda_-)}}{2\pi \lambda} & 
(\lambda_- \le \lambda \le \lambda_+),\\
0 & (\text{otherwise}),
\end{cases}
\label{eq:Marcenko}
\end{equation}
where $\lambda_{\pm} = \left(1 \pm \sqrt{N/L}\right)^2$
\cite{Laloux1999PhysRevLett,Plerou1999PhysRevLett,Macmahon2015PhysRevX}
(shown in the black lines in Fig.~\ref{fig:eigs random}). The results are qualitatively the same for a larger random correlation matrix with $N=500$ and $L=1,000$ (Fig.~\ref{fig:eigs random}(b)). Therefore, when random correlation matrices are input, the present configuration model behaves similarly to the existing null models $\langle \rho^{\rm MG2} \rangle$
and $\langle \rho^{\rm MG3} \rangle$.

Then, we turn to a random correlation matrix with community structure. By adapting the benchmark models used in Ref.~\cite{Macmahon2015PhysRevX}, we construct a random correlation matrix with four non-overlapping communities as follows. We set $N=500$ and $L=1,000$. We assume that the signal on the $i$th node ($1\le i\le N$) at time $t$ ($1\le t\le L$) is given by
$x_{it} = \mu \alpha(t) + \nu \beta_i(t) + \gamma_c(t)$, where $\alpha(t)$, $\beta_i(t)$ and $\gamma_c(t)$ for each $i$ ($1\le i\le N$), $c$ ($1\le c\le 4$) and $t$ ($1\le t\le L$)
are independent normal variables with mean zero and standard deviation 1. Signal $\alpha(t)$ represents the global signal, $\beta_i(t)$ represents local noise, $\gamma_c(t)$ corresponds to the signal for each community, $\mu$ represents the strength of the global signal, and $\nu$ represents the strength of the local noise. We set $\mu=0.4$ and $\nu=0.8$ and assume that 
$c=1$ for $1\le i\le 50$, $c=2$ for $51\le i\le 150$, $c=3$ for $151\le i\le 300$ and $c=4$ for $301\le i\le 500$, thus generating four communities of size 50, 100, 150 and 200.

The distribution of eigenvalues for a sample correlation matrix with four communities is shown in skyblue in Fig.~\ref{fig:eigs random}(c). The distribution is composed of a bulk of eigenvalues and four large eigenvalues that do not belong to the bulk. The bulk part of the distribution does not resemble the Marcenko-Pastur distribution, whereas the eigenvalues are not considerably larger than $\lambda_+$, i.e., the largest value for the Marcenko-Pastur distribution. The four largest eigenvalues correspond to the four planted communities. The eigenvalue distribution for a sample correlation matrix generated by the estimated configuration model is shown by the red lines in Fig.~\ref{fig:eigs random}(c). It consists of a bulk part and a single large eigenvalue. The bulk part deviates from the Marcenko-Pastur distribution. However, it is closer to the Marcenko-Pastur distribution than the bulk part of the eigenvalue distribution for the original correlation matrix with four communities (shown in skyblue in Fig.~\ref{fig:eigs random}(c)) is. Note that the three additional eigenmodes corresponding to the communities are filtered out by the configuration model (shown in red in Fig.~\ref{fig:eigs random}(c)). The present configuration model is expected to be suitable as a null model for community detection (Section~\ref{sec:community}) because the model filters out the singular eigenmodes encoding the community structure.

\begin{figure}
\begin{center}
% Edits by Kojaku
\includegraphics[width=\hsize]{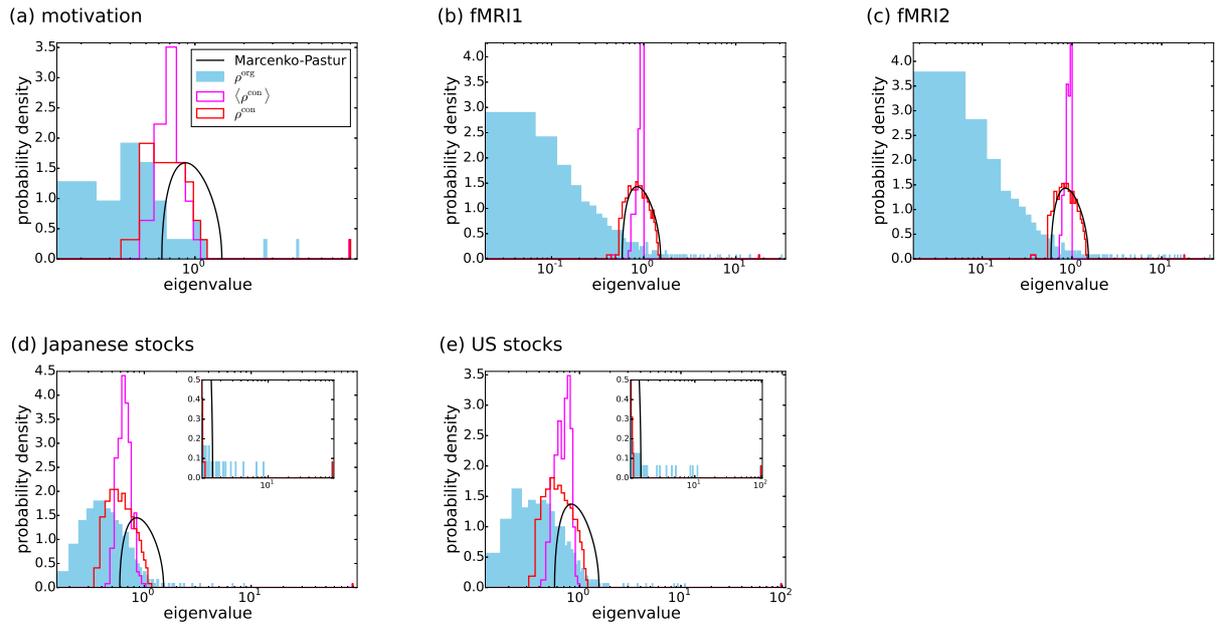}
\caption{Density of the eigenvalues of the empirical correlation matrix, denoted by $\rho^{\rm org}$, and the corresponding configuration model, denoted by $\rho^{\rm con}$. The black solid lines represent the Marcenko-Pastur distribution. (a) Motivation. (b) fMRI1. (c) fMRI2. (d) Japanese stocks. (e) US stocks. The insets in (d) and (e) are magnifications of the main figures.}
\label{fig:eigs empirical}
\end{center}
\end{figure}

Next, for the five empirical correlation matrices, we compared the distribution of eigenvalues between the original correlation matrix and a sample correlation matrix generated by the estimated configuration model. The results are shown in Fig.~\ref{fig:eigs empirical}. The figures suggest that, for the fMRI data, the configuration model produces a distribution of eigenvalues that is almost the same as the Marcenko-Pastur distribution except for one eigenmode whose eigenvalue is much larger than $\lambda_+$ (red lines in Figs.~\ref{fig:eigs empirical}(b) and (c)). The mode with the largest eigenvalue, which we call the dominant mode (also called the market mode in the literature \cite{Macmahon2015PhysRevX}), corresponds to the conservation of the node's strength, as we will examine in the next section. Although the largest eigenvalue of $\rho^{\rm con}$ is different from that of the original correlation matrix, $\rho^{\rm org}$, due to randomness of $\rho^{\rm con}$ and possibly for other reasons, the eigenvalue distribution for $\rho^{\rm con}$ is similar to that for
$\langle \rho^{\rm MG3} \rangle$.

Because the present configuration model is a Wishart distribution of covariance matrices, we have access to its expectation with respect to $p(X)$,
which is equal to $\Sigma$ for any $L$. We convert $\Sigma$
to the correlation matrix to denote it by
$\langle \rho^{\rm con}\rangle$, where $\langle \cdot\rangle$ represents the expectation. Correlation matrix $\langle \rho^{\rm con}\rangle$ is approximately the expectation of the sample correlation matrix, $\rho^{\rm con}$.
Note that $\langle \rho^{\rm con}\rangle$ is equal to any sample correlation matrix, 
$\rho^{\rm con}$, in the limit $L\to\infty$. The eigenvalue distribution for
$\langle \rho^{\rm con}\rangle$ is shown by the magenta lines in Fig.~\ref{fig:eigs empirical}. If the distribution followed the combination of a single dominant eigenvalue and the Marcenko-Pastur distribution, Eq.~\eqref{eq:Marcenko} suggests that the bulk part would follow the delta function located
at $\lambda = 1$ because $\langle \rho^{\rm con}\rangle$ corresponds to the limit $L\to\infty$. However,
the figure suggests that this is not the case. The eigenvalue distribution of $\langle \rho^{\rm con}\rangle$ is composed of a noisy part with a finite width and a dominant mode.

For the motivation data (Fig.~\ref{fig:eigs empirical}(a)) and the financial data
(Figs.~\ref{fig:eigs empirical}(d) and \ref{fig:eigs empirical}(e)), the bulk part of the eigenvalue distribution for
$\rho^{\rm con}$ deviates from the Marcenko-Pastur distribution. However, it is closer to the Marcenko-Pastur distribution than the bulk part of the eigenvalue distribution for the original correlation matrix is. In addition, $\rho^{\rm con}$ has a single dominant eigenvalue that is much larger than the other eigenvalues. These observations also apply to $\langle \rho^{\rm con}\rangle$. Therefore, for the motivation and financial data, the present configuration model filters the input correlation matrix to produce a correlation matrix that is qualitatively, although not quantitatively, similar to $\langle \rho^{\rm MG3} \rangle$.

\subsection{Strength of each node}

In this section, we compare the strength of each node between the empirical correlation matrices and those generated by different models. 

\begin{figure}
\begin{center}
% Edits by Kojaku
\includegraphics[width=\hsize]{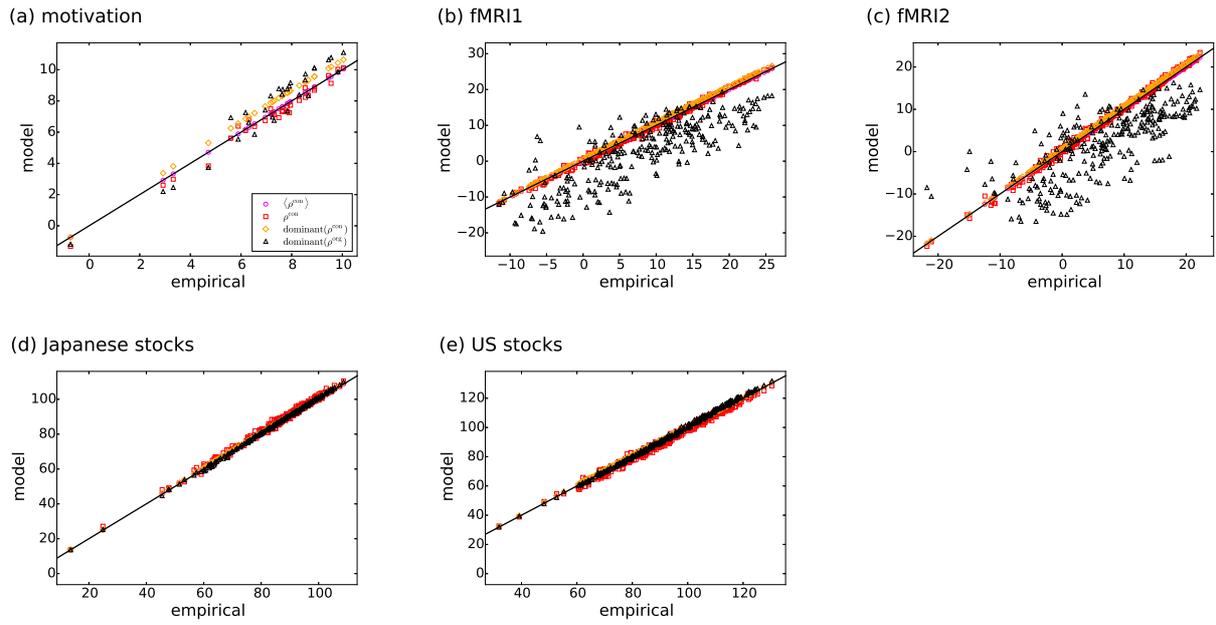}
\caption{Comparison of the node strength between the original correlation matrix, the configuration model and the correlation matrices constructed from the dominant mode. (a) Motivation. (b) fMRI1. (c) fMRI2. (d) Japanese stocks. (e) US stocks.}
\label{fig:s scatter config}
\end{center}
\end{figure}

The strength of each node, defined by $s_i = \sum_{j=1; j\neq i}^N \rho_{ij}$, is compared between each of the empirical correlation matrices, $\rho^{\rm org}$, and the corresponding configuration model in Fig.~\ref{fig:s scatter config}. For all the empirical correlation matrices, $\langle \rho^{\rm con}\rangle$ almost perfectly reproduces the strength of each node in $\rho^{\rm org}$, corroborating the validity of our gradient descent algorithm (shown by the circles in Fig.~\ref{fig:s scatter config}). A sample correlation matrix $\rho^{\rm con}$ generated by the configuration model produces node strengths that carry some fluctuations around the correct values (squares in Fig.~\ref{fig:s scatter config}). Because the standard deviation of each entry of $\rho^{\rm con}$ is proportional to 
$L^{-1/2}$ (Section~\ref{sec:model}),
the fluctuation is generally small for data with a large $L$ value.

The eigenvalue distribution for the configuration model is characterised by a dominant mode and the $N-1$ eigenvalues that constitute a bulk that resembles the Marcenko-Pastur distribution to different extents depending on the data (Fig.~\ref{fig:eigs empirical}). To examine the relationship between the largest eigenvalue and the conservation of the node's strength, we filter the expected correlation matrix generated by the present configuration model, $\langle\rho^{\rm con}\rangle$, by only keeping the dominant eigenmode. In other words, we calculate matrix $\lambda_1 \bm u_{(1)} \bm u_{(1)}^{\top}$, where $\lambda_1$ is the largest eigenvalue of $\langle\rho^{\rm con}\rangle$ and
$\bm u_{(1)}$ is the corresponding normalised column eigenvector. Then, we compute the node's strength for $\lambda_1 \bm u_{(1)} \bm u_{(1)}^{\top}$. It should be noted that, although $\lambda_1 \bm u_{(1)} \bm u_{(1)}^{\top}$ is not a correlation matrix because its diagonal elements are not equal to unity in general, the diagonal elements are not used in the calculation of the node's strength such that the node's strength is well defined \cite{Macmahon2015PhysRevX}.

The node strength for $\lambda_1 \bm u_{(1)} \bm u_{(1)}^{\top}$ is plotted against 
that for the original correlation matrix, $\rho^{\rm org}$, by the diamonds in Fig.~\ref{fig:s scatter config}. Despite a slight overestimation, $\lambda_1 \bm u_{(1)} \bm u_{(1)}^{\top}$ reproduces the node's strength for the original correlation matrix with a high accuracy. Therefore, our configuration model roughly retains the dominant mode of the original correlation matrix to conserve the node's strength and produce the other $N-1$ random modes whose eigenvalue distribution approximates the Marcenko-Pastur distribution to different extents. However, differently from a previous null model, $\langle \rho^{\rm MG3} \rangle$, that exactly preserves the dominant mode of the input correlation matrix, the dominant mode of the present configuration model is not the same as that of the original correlation matrix. This fact is evinced by the difference in the position between the rightmost skyblue versus magenta bars in each panel of Fig.~\ref{fig:eigs empirical}.

To examine the relationship between the node strength and the dominant mode of the
empirical correlation matrices, we calculated matrix
$\lambda_1 \bm u_{(1)} \bm u_{(1)}^{\top}$ from the original correlation matrix and plotted its node strength against that of the original correlation matrix $\rho^{\rm org}$
by the triangles in Fig.~\ref{fig:s scatter config}. Note that this particular analysis does not have to do with any null model including the present configuration model. For the financial data, the dominant mode of $\rho^{\rm org}$ explains the strength of each node with a high accuracy (Figs.~\ref{fig:s scatter config}(d) and \ref{fig:s scatter config}(e)). This is presumably because the dominant eigenvalue is much larger than the other $N-1$ eigenvalues for these correlation matrices, which is a robust observation for financial time series data \cite{Laloux1999PhysRevLett,Plerou1999PhysRevLett,Macmahon2015PhysRevX}.
For the motivation and fMRI data, for which the dominant eigenvalue is not relatively large as compared to the case of the financial data, we also find a similar agreement between the dominant mode and the node strength albeit with a lower accuracy (Figs.~\ref{fig:s scatter config}(a)--(c)). 

We conclude that the dominant mode represents the sequence of node strength if the largest eigenvalue is far from the other eigenvalues of the correlation matrix. To our numerical effort, this condition holds true for some empirical correlation matrices and all correlation matrices obtained from the configuration model.

\begin{figure}
\begin{center}
% Edits by Kojaku
\includegraphics[width=\hsize]{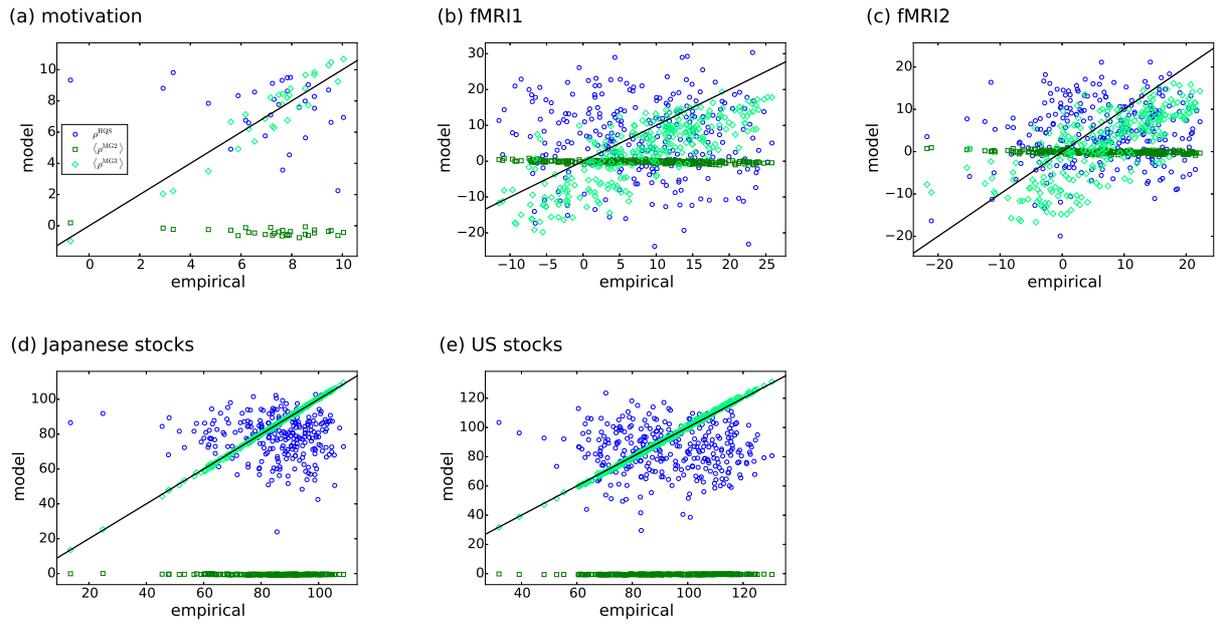}
\caption{Comparison of the node strength between the original correlation matrix, the H-Q-S model and the models based on random matrix theory. (a) Motivation. (b) fMRI1. (c) fMRI2. (d) Japanese stocks. (e) US stocks.}
\label{fig:s scatter HQS-MG}
\end{center}
\end{figure}

Next, we examine the same relationship between the empirical correlation matrices and three other models of correlation matrix. The first correlation matrix is a covariance matrix generated by the H-Q-S algorithm (Appendix~\ref{app:HQS}), which is then converted to the correlation matrix. We denote this correlation matrix by $\rho^{\rm HQS}$. The other two correlation matrices are 
derived from random matrix theory, i.e., $\langle \rho^{\rm MG2} \rangle$ and $\langle \rho^{\rm MG3}\rangle$.
The strength of each node is compared between the empirical correlation matrices, $\rho^{\rm org}$, and the three models in Fig.~\ref{fig:s scatter HQS-MG}. 
Correlation matrix $\langle \rho^{\rm MG3} \rangle$ reproduces the node strength with a high accuracy for the financial data
(diamonds in Figs.~\ref{fig:s scatter HQS-MG}(d) and \ref{fig:s scatter HQS-MG}(e)).
This result is consistent with the observation that the dominant mode reproduces
the node strength (Figs.~\ref{fig:s scatter config}(d) and \ref{fig:s scatter config}(e)).
We obtain qualitatively the same results for the other data sets although
the association between the empirical correlation matrix and
$\langle \rho^{\rm MG3} \rangle$ in terms of the node strength is weaker
(diamonds in Figs.~\ref{fig:s scatter HQS-MG}(a)--(c)).

Correlation matrices $\rho^{\rm HQS}$ and $\langle \rho^{\rm MG2}\rangle$ do not produce heterogeneous distributions of the strength across different nodes
(circles and squares in Fig.~\ref{fig:s scatter HQS-MG}).
In particular, the node strength for $\langle \rho^{\rm MG2}\rangle$ is close to zero for all nodes. They can be regarded as correlation-matrix counterparts of the Erd\H{o}s-R\'{e}nyi random graph for conventional networks, which do not conserve each node's degree.

\subsection{Distribution of off-diagonal elements}

\begin{figure}
\begin{center}
% Edits by Kojaku
\includegraphics[width=\hsize]{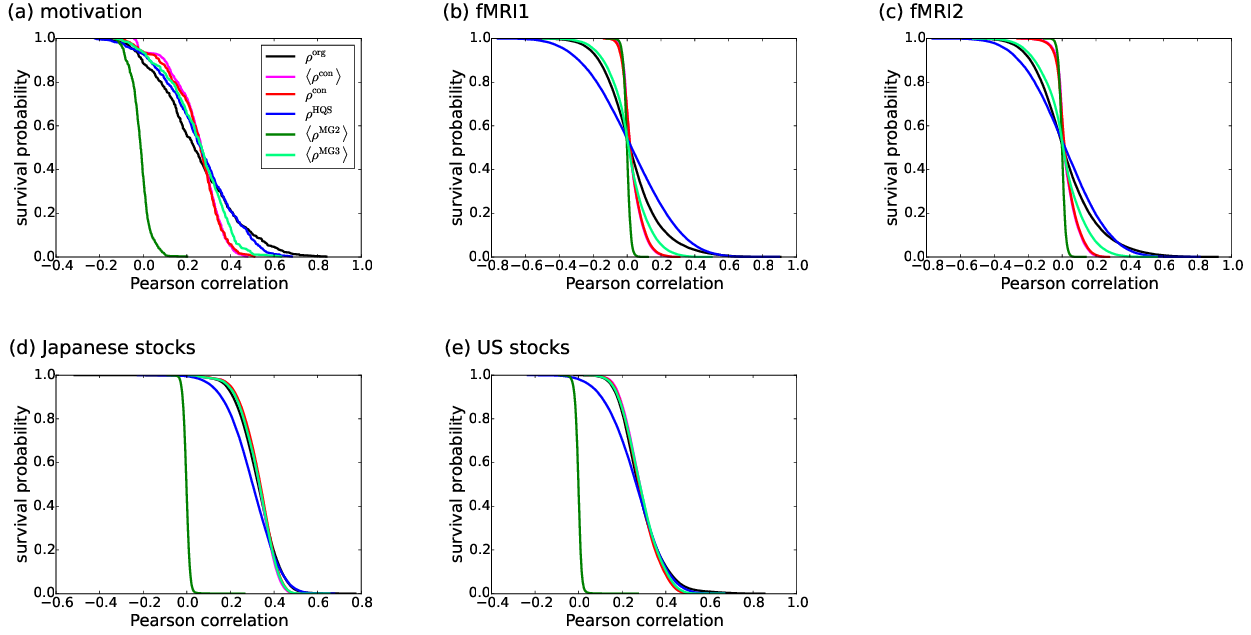}
\caption{Survival probability of the values of the off-diagonal elements in the correlation matrix. (a) Motivation. (b) fMRI1. (c) fMRI2. (d) Japanese stocks. (e) US stocks.}
\label{fig:off-diagonal dist}
\end{center}
\end{figure}

The survival probability of the off-diagonal elements of the correlation matrix (i.e., Pearson correlation values between pairs of nodes) is compared between the empirical data and the models in Fig.~\ref{fig:off-diagonal dist} for each data set. The expectation of the configuration model, $\langle \rho^{\rm con}\rangle$, produces distributions of the off-diagonal elements moderately close to the empirical distributions. As expected,
$\rho^{\rm con}$ produces somewhat noisier distributions.
Correlation matrix $\langle \rho^{\rm MG3} \rangle$ beats our configuration model
(i.e., $\langle \rho^{\rm con}\rangle$ and $\rho^{\rm con}$)
in approximating the empirical distribution. The H-Q-S model, $\rho^{\rm HQS}$, also produces distributions roughly close to the empirical ones, which is consistent with the previous results \cite{Hirschberger2007EurJOperRes,Zalesky2012Neuroimage,Hosseini2013PlosOne}. The distributions derived from $\langle \rho^{\rm MG2}\rangle$ are far from the empirical distributions.

\subsection{Clustering coefficient\label{sec:clustering}}

Clustering coefficients measure abundance of connected triangles in networks. For conventional networks, the cluster coefficients in empirical networks are much larger than in the configuration model in many cases \cite{Newman2010book}. For correlation matrix data, one can construct a conventional weighted network by using the Pearson correlation value as the edge weight or an conventional unweighted network by thresholding on the edge weight. In both cases, the clustering coefficient tends to be inflated due to the presence of an indirect path (correlation between nodes $v_1$ and $v_2$ and that between $v_1$ and $v_3$ implies correlation between $v_2$ and $v_3$) \cite{Zalesky2012Neuroimage,Masuda2017submitted}.
The H-Q-S model was shown to mitigate the effect of indirect paths on statistically measuring clustering coefficients
\cite{Zalesky2012Neuroimage,Hosseini2013PlosOne}. In this section, we compare the impact of different null models on the statistical significance of clustering coefficients in empirical correlation matrices.
We use three models as null models, i.e., an algorithm that generates correlation matrices by assuming relatively long white-noise signals independent across different nodes, which we call the white-noise model (Appendix~\ref{app:white}), the H-Q-S model and our configuration model. We do not use $\langle \rho^{\rm MG2}\rangle$ or $\langle \rho^{\rm MG3}\rangle$ because they are not designed to produce random samples of correlation matrices, which are necessary for calculating the statistical significance of the clustering coefficients or other indices.

Because various measurements of unweighted correlation networks depend on the threshold value
\cite{Zalesky2012Neuroimage,Garrison2015Neuroimage,Jalili2016SciRep}, we use two types of clustering coefficients that do not require thresholding. The first clustering coefficient is a weighted clustering coefficient \cite{Onnela2005PhysRevE}, denoted by
$C^{\rm wei,O}$ (Appendix~\ref{app:C^{wei,O}}).
The second clustering coefficient, denoted by $C^{\rm cor,M}$, is the one based on partial mutual information, which we recently proposed \cite{Masuda2017submitted} (Appendix~\ref{app:C^{cor,M}}).

For each empirical correlation matrix and each null model, we generate $10^3$ correlation matrices, calculate the clustering coefficient (i.e., $C^{\rm wei,O}$ or $C^{\rm cor,M}$) for each of the generated correlation matrices and calculate the sample mean and standard deviation of the clustering coefficient, denoted by $\tilde{\mu}$ and $\tilde{\sigma}$, respectively. 
The $Z$ score is given by $(C^{\rm org} - \tilde{\mu})/\tilde{\sigma}$, where $C^{\rm org}$ is the clustering coefficient for the original correlation matrix.
By assuming that the clustering coefficient for the null model obeys a normal distribution, we translate the $Z$ score to the $P$ value based on the two-tailed test.

For the five empirical correlation matrices, the values of the clustering coefficients and the statistical results are shown in Table~\ref{tab:C results}. 
Both $C^{\rm wei,O}$ and $C^{\rm cor,M}$ for all the empirical networks
are significantly larger than the values for the white-noise null model. This result is consistent with common knowledge that many empirical networks have high clustering \cite{Newman2010book}, including the case of weighted networks  \cite{Saramaki2007PhysRevE}. However, the same result does not hold true for the other two null models. Relative to the present configuration model, $\rho^{\rm con}$,
clustering coefficient $C^{\rm wei,O}$ is significantly small for all the five empirical correlation matrices. In contrast, $C^{\rm cor,M}$ for all the empirical correlation matrices is larger than that for $\rho^{\rm con}$,
including the case of insignificant results (i.e., Japanese stock market data). With the H-Q-S null model, the results vary across both the empirical correlation matrix and the type of clustering coefficient.

In sum, the empirical correlation matrices do not necessarily show high clustering coefficients when the H-Q-S model or the present configuration model is used as the reference. In addition, the selection of the null model (i.e., the H-Q-S versus configuration model) may even qualitatively change the statistical results.

\subsection{Community detection\label{sec:community}}

Various conventional networks are organised into communities, i.e., sets of nodes such that the edges are dense within a community and relatively sparse across different communities \cite{Fortunato2010PhysRep}.
In this section, we apply our configuration model to community detection in correlation matrices. A naive application of community detection algorithms designed for conventional weighted networks to correlation matrix data would yield biased results. This observation led to development of community-detection algorithms tailored to correlation matrices with appropriate null models \cite{Macmahon2015PhysRevX}.
We compare community detection when the null model is either $\langle \rho^{\rm con} \rangle$, the expectation of $\rho^{\rm HQS}$ denoted by
$\langle \rho^{\rm HQS}\rangle$, $\langle \rho^{\rm MG2}\rangle$, $\langle \rho^{\rm MG3}\rangle$, or the identity matrix denoted by $\langle \rho^{\rm MG1}\rangle$. All the off-diagonal values of $\langle \rho^{\rm HQS}\rangle$ are equal (Appendix~\ref{app:HQS}).
Correlation matrix $\langle \rho^{\rm MG1}\rangle$ assumes the absence of correlation between any pair of nodes.
Note that $\langle \rho^{\rm MG1}\rangle$, $\langle \rho^{\rm MG2}\rangle$ and $\langle \rho^{\rm MG3}\rangle$ have been used for community detection in correlation matrices \cite{Macmahon2015PhysRevX}. 

We maximise the modularity given by \cite{Macmahon2015PhysRevX}
\begin{equation}
Q = \frac{1}{C_{\rm norm}} \sum_{i, j=1}^N \left( \rho_{ij} - \langle \rho_{ij} \rangle \right)
\delta (g_i, g_j),
\label{eq:Q}
\end{equation}
where 
$C_{\rm norm} = \sum_{i, j=1}^N \rho_{ij}$ is a normalisation constant,
$\langle \rho\rangle$ is a null model of the correlation matrix relative to which community structure is detected,
$\delta$ is the Kronecker delta, and $g_i$ is the community to which node $i$ belongs. We use the Louvain algorithm \cite{Blondel2008JStatMech} to maximise 
$Q$.
% To maximise $Q$ with null models $\langle \rho^{\rm MG1} \rangle$, $\langle \rho^{\rm MG2}\rangle$ and $\langle \rho^{\rm MG3}\rangle$, we use the code provided in Ref.~\cite{Macmahon2015PhysRevX}.

To assess the statistical significance of the detected community structure, we maximise $Q$ for randomised correlation matrices as well as for the given correlation matrix. When the null model is our configuration model, we generated random samples $\rho^{\rm con}$ to calculate the $Z$ score and $P$ value. When the null model is $\langle \rho^{\rm HQS}\rangle$, we generated random samples $\rho^{\rm HQS}$ from the H-Q-S model. Because $\langle \rho^{\rm MG1} \rangle$, $\langle \rho^{\rm MG2}\rangle$ and $\langle \rho^{\rm MG3}\rangle$ are null models that do not generate sample correlation matrices, we generated random samples from the H-Q-S model (i.e., $\rho^{\rm HQS}$) for these null models. In each case, we generated $10^3$ random correlation matrices to calculate the $Z$ score and the $P$ value.

First, we start by using $\langle\rho^{\rm con}\rangle$ as the input correlation matrix rather than the null model. Correlation matrix $\langle\rho^{\rm con}\rangle$ is considered to lack community structure because it is maximally random in terms of the entropy under the constraint on the strength of each node. Because the modularity value would be trivially insignificant if $\langle\rho^{\rm con}\rangle$ is used as the null model, we maximised the modularity with the other four null models, i.e.,
$\langle\rho^{\rm HQS}\rangle$, $\langle \rho^{\rm MG1} \rangle$, $\langle \rho^{\rm MG2}\rangle$ and $\langle \rho^{\rm MG3}\rangle$.
The optimized $Q$ values and the statistical results for the different empirical networks are shown in Table~\ref{tab:comm config}. The modularity with the H-Q-S null model has detected significant community structure in the configuration-model correlation matrix (i.e., $\langle\rho^{\rm con}\rangle$) for all the data sets. Similarly, the modularity with 
the $\langle \rho^{\rm MG1} \rangle$ null model yields significant community structure in two cases with $N=264$. Therefore, we conclude that these two null models are not suitable for community detection. In contrast, modularity with the $\langle \rho^{\rm MG2}\rangle$ or $\langle \rho^{\rm MG3}\rangle$ null model does not find significant community structure, except for $\langle \rho^{\rm MG2}\rangle$ with the motivation data, which is a small data set ($N=30$). Therefore,
$\langle \rho^{\rm MG2}\rangle$ and $\langle \rho^{\rm MG3}\rangle$ seem to be reasonable null models for community detection \cite{Macmahon2015PhysRevX}.

Therefore, we focus on community structure of the empirical correlation matrices obtained by maximising $Q$ combined with either the $\langle\rho^{\rm con}\rangle$, $\langle \rho^{\rm MG2}\rangle$ or $\langle \rho^{\rm MG3}\rangle$ null model.
The maximised modularity values and statistical results for the empirical data are shown in Table~\ref{tab:comm empirical}. The maximised modularity is insignificant for all the five empirical correlation matrices when the null model is $\langle \rho^{\rm MG3}\rangle$. The modularity is significant for all but the fMRI1 data when the null model is
$\langle \rho^{\rm MG2}\rangle$. It should be noted that, with the combination of $\langle \rho^{\rm MG2}\rangle$ and either the Japanese or US stock data, the modularity value is almost equal to 1 for any randomised correlation matrices. This is because the magnitude of the eigenvalues whose corresponding eigenmodes are preserved
in $\langle \rho^{\rm MG2}\rangle$ is much smaller than the dominant eigenvalue. Then, $\langle \rho^{\rm MG2}\rangle$ is approximately a zero matrix, which makes the second term on the right-hand side of Eq.~\eqref{eq:Q} negligible. Furthermore, modularity maximisation has only detected a single community (i.e., no partition into different communities), which makes the summation on the right-hand side of Eq.~\eqref{eq:Q} almost equal to $C_{\rm norm}$, yielding $Q\approx 1$. With the configuration null model, the modularity is significant in all cases, presumably because the value of $L$ is large and fluctuations of the modularity for samples generated by the estimated configuration model are small.

\begin{figure}
\centering
\includegraphics[width=1.0\hsize]{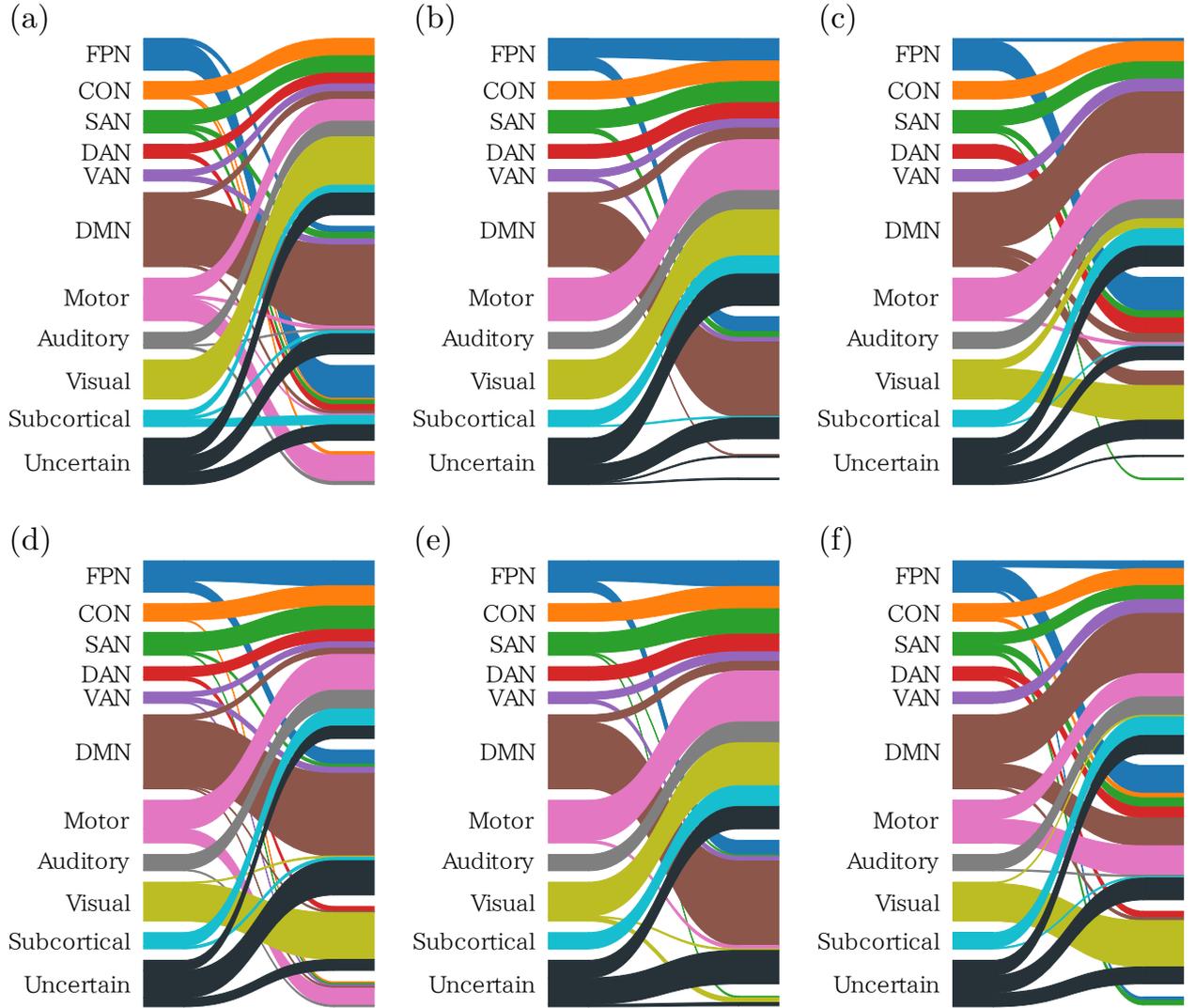}
\caption{Community structure for the fMRI data. (a)--(c): fMRI1. (d)--(f): fMRI2.
The null model is $\langle \rho^{\rm con}\rangle$ in (a) and (d),
$\langle \rho^{\rm MG2}\rangle$ in (b) and (e), and
$\langle \rho^{\rm MG3}\rangle$ in (c) and (f). In each panel, the nodes are vertically stacked. The labels to the left indicate the name of the brain system to which each node belongs. 
FPN: fronto-parietal network, CON: cingulo-opercular network,
SAN: salience network, DAN: dorsal attention network, VAN: ventral attention network,
DMN: default mode network. Uncertain indicates that the node does not belong to a particular brain system. A bundle to the right in each panel represents a community detected by modularity maximisation. For example, in (a), there are four communities, the smallest one of which shown to the bottom mostly consists of the nodes in the motor network.}
\label{fig:alluvial}
\end{figure}

Because the motivation data set is small and the modularity values for the original correlation matrices are small for the stock market data, we focus on the fMRI data in the remainder of this section.
In the present fMRI data, each node is assigned with a biologically determined label representing estimated functions of the node \cite{Power2011Neuron}. The relationship between the detected community structure and the biological label of the node is shown in Fig.~\ref{fig:alluvial}.

To assess the extent to which the detected communities are consistent with the biological label of the node, we compute the probability that two nodes with the same label belong to the same community. We denote this probability by $P^{\rm emp}$.
Because $P^{\rm emp}$ would be large when there are a small number of communities, we normalise $P^{\rm emp}$ by the probability in the case of the completely random assignment of nodes to a label, which we denote by $P^{\rm rand}$.
We obtain
$P^{\rm rand} = \left[N(N-1)/2\right]^{-1} \times \sum_{c=1} ^{n_{\rm comm}} N_c (N_c -1)/2$,
where $n_{\rm comm}$ is the number of communities and $N_c$ is the number of nodes in the $c$th community. The values of 
$P^{\rm emp} - P^{\rm rand}$ and $P^{\rm emp}/P^{\rm rand}$ for the two fMRI data sets and the three null models are shown in the top half of Table~\ref{tab:P^{emp}}.
For both normalised measures of the consistency between the nodal label and community structure, i.e., $P^{\rm emp} - P^{\rm rand}$ and $P^{\rm emp}/P^{\rm rand}$,
the configuration null model realises a larger value than the $\langle \rho^{\rm MG2}\rangle$ and $\langle \rho^{\rm MG3}\rangle$ null models do. The results remain the same when the nodes having label ``Uncertain'' are removed before $P^{\rm emp}$ and $P^{\rm rand}$ are calculated (the bottom half of Table~\ref{tab:P^{emp}}). We conclude that, for the present data set, 
our configuration model produces community structure that is more consistent with the biological label than the $\langle \rho^{\rm MG2}\rangle$ and $\langle \rho^{\rm MG3}\rangle$ null models do.

However, a visual inspection of Fig.~\ref{fig:alluvial} suggests that the $\langle \rho^{\rm MG2}\rangle$ null model realises community structure that is more consistent between the two participants than the other two null models do
(Figs.~\ref{fig:alluvial}(b) and \ref{fig:alluvial}(e) as compared to Figs.~\ref{fig:alluvial}(a), \ref{fig:alluvial}(c), \ref{fig:alluvial}(d) and \ref{fig:alluvial}(f)).
To examine this point, we measure
the Jaccard index between the community structure detected for fMRI1 and that for fMRI2.
The Jaccard index is defined by
$\sum_{i=1}^N \sum_{j=1}^{i-1} \delta(g^{(1)}_i,g^{(1)}_j)\delta(g^{(2)}_i,g^{(2)}_j)$
$\big/$
$\sum_{i=1}^N \sum_{j=1}^{i-1} \left[ \delta(g^{(1)}_i,g^{(1)}_j) + \delta(g^{(2)}_i,g^{(2)}_j) - \delta(g^{(1)}_i,g^{(1)}_j)\delta(g^{(2)}_i,g^{(2)}_j) \right]$, where
$g_i ^{(1)}$ and $g_i ^{(2)}$ are the communities to which node $i$ belongs in the fMRI1 and fMRI2 data, respectively. We have found that the Jaccard index is larger (therefore, the two community structures are more similar) for the 
the $\langle \rho^{\rm MG2}\rangle$ null model ($= 0.567$) than
the $\langle \rho^{\rm con}\rangle$ ($=0.313$) and $\langle \rho^{\rm MG3}\rangle$ 
($=0.418$) null models.

\section{Discussion}

We proposed a configuration model for correlation matrices that preserves the expected strength of each node. We illustrate applications of the present model with clustering coefficients and community detection. Being a configuration model, the present model will find applications in measurements and algorithms for correlation and covariance matrices where comparison between the original matrix and reference matrices (i.e., null models) will be important. Judging from similar situations for conventional networks, we expect application of the present paper in, for example, different algorithms of community detection \cite{Fortunato2010PhysRep},
network motifs \cite{Milo2002Science} and
detection of core-periphery structure \cite{Kojaku2017arxiv}.

A correlation matrix can be regarded as a weighted network. 
Several configuration models including those based on the maximum entropy principle have been proposed for weighted networks
\cite{Opsahl2008PhysRevLett,Serrano2008PhysRevE,Garlaschelli2009PhysRevLett,Squartini2011NewJPhys,Mastrandrea2014NewJPhys,Almog2015NewJPhys,Cimini2015PhysRevE,Squartini2015NewJPhys}. However, differently from the present configuration model,
these previous models do not conserve positive semidefiniteness, which any correlation or covariance matrix must satisfy. In addition, our configuration model allows negative entries, whereas the previous models exclude negative edge weights; correlation or covariance matrices generally have negative entries. Therefore, the maximum entropy models for conventional weighted networks \cite{Garlaschelli2009PhysRevLett,Squartini2011NewJPhys,Mastrandrea2014NewJPhys,Almog2015NewJPhys,Cimini2015PhysRevE,Squartini2015NewJPhys} and our model are different although they share the maximum entropy principle.

As a separate issue, constructing a weighted configuration model for conventional weighted networks is inherently difficult due to structural constraints imposed by the topology of the corresponding unweighted network \cite{Serrano2006PhysRevE-weighted}. With our configuration model, we evaded this difficulty by not imposing an unweighted network topology in the estimated correlation or covariance matrix. 

Correlation matrix $\langle \rho^{\rm MG3}\rangle$ derived from random matrix theory \cite{Macmahon2015PhysRevX} is similar to the present configuration model in the sense that $\langle \rho^{\rm MG3}\rangle$ fairly accurately produced the node strength for the motivation data and financial data (Figs.~\ref{fig:s scatter HQS-MG}(a), \ref{fig:s scatter HQS-MG}(d) and \ref{fig:s scatter HQS-MG}(e)). For the fMRI data, it also explained the node strength for the fMRI data albeit to a lesser extent (Figs.~\ref{fig:s scatter HQS-MG}(b) and \ref{fig:s scatter HQS-MG}(c)). This is because $\langle \rho^{\rm MG3}\rangle$ preserves the dominant eigenmodes of the original matrix by definition. The dominant eigenmode is strongly correlated with the node's strength (Fig.~\ref{fig:s scatter config}).
In conventional unweighted networks, the eigenvector corresponding to the largest eigenvalue, called the eigenvector centrality \cite{Bonacich1972JMathSociol}, is often correlated with the node's degree \cite{Newman2010book,Estrada2012book}. Also from this point of view, it is natural that the dominant mode approximately conserves the node's strength in various correlation matrices.
Another similarity between the present configuration model and $\langle \rho^{\rm MG3}\rangle$ is in the eigenvalue distribution (Figs.~\ref{fig:eigs random} and \ref{fig:eigs empirical}). The configuration model also roughly preserves the dominant eigenmode. The configuration model does not perfectly preserve the bulk of the eigenvalue distribution corresponding to that of random correlation matrices (i.e., those falling in the support of the Marcenko-Pastur distribution), differently from $\langle \rho^{\rm MG3}\rangle$. Nevertheless, the configuration model shifts the bulk part of the eigenvalue distribution closer to the Marcenko-Pastur distribution.
A difference between the present configuration model and $\langle \rho^{\rm MG3}\rangle$ is that the former generally preserves the rank of the given correlation matrix, whereas the latter has a smaller rank owing to the elimination of some eigenmodes.
Another difference is that
$\langle \rho^{\rm MG3}\rangle$ requires the length of the data (e.g., time series) based on which the correlation matrix is calculated, $L$, whereas the configuration model does not. The configuration model does need $L$ to produce sample correlation matrices (i.e., $\rho^{\rm con}$). However, it can be used in another mode, which is the expectation of the produced correlation matrices (i.e., $\langle \rho^{\rm con}\rangle$). In fact, we used $\langle \rho^{\rm con}\rangle$ for community detection
(Section~\ref{sec:community}). This usage does not require the $L$ value.

In sum, both $\langle \rho^{\rm MG3}\rangle$ and the present configuration model can be regarded as configuration models for correlation matrices. 
They provide different methods to filter noise in correlation matrices.
In contrast, the H-Q-S algorithm and $\langle \rho^{\rm MG2}\rangle$ disregard the node's heterogeneity. Therefore, they are regarded as counterparts of the Erd\H{o}s-R\'{e}nyi random graph for correlation matrices.

An important limitation of the proposed algorithm is scalability. The gradient descent algorithm used in the present paper is slow because, in our experience, we have to make the learning rate (i.e., $\epsilon$ in Eqs.~\eqref{eq:gradient descent alpha_i} and \eqref{eq:gradient descent beta_i}) small for the algorithm to converge. Therefore, the largest correlation matrix that we used in the present paper was of size $N=500$. 
Alternatively, one can formulate a multidimensional root finding problem with unknowns $\alpha_i$ and $\beta_i$ ($1\le i\le N$) (Appendix~\ref{app:root finding}). However, we could not find the roots, which may be because of strong nonlinearity inherent in the set of the equations.
The corresponding optimisation problem does not seem to be convex.
The entropy probably has a rough landscape as a function of $\alpha_i$
and $\beta_i$. Up to our numerical efforts, we found that the landscape was even more rough when we fed covariance matrices rather than correlation matrices to our algorithm. Understanding this issue and devising more efficient algorithms are left for future work.

\section*{Acknowledgments}

We thank Diego Garlaschelli for valuable discussion.
We thank Koh Murayama for providing the academic motivation data used in the present paper.
We thank Takahiro Ezaki for calculating the correlation matrices for the stock market data.
The fMRI data were provided in part by the Human Connectome Project, Washington University--Minn Consortium (Principal Investigators: David Van Essen and Kamil Ugurbil; 1U54MH091657) funded by the 16 NIH Institutes and Centers that support the National Institute of Health (NIH) Blueprint for Neuroscience Research, and by the McDonnell Center for Systems Neuroscience at Washington University.
N.M. acknowledges the support provided through Japan Science and Technology Agency (JST) CREST Grant No. JPMJCR1304 and the JST ERATO Grant No. JPMJER1201, Japan.

% I thank Sidney Redner for valuable discussion throughout the current work.
% I also thank Yuni Iwamasa and Taro Takaguchi for careful reading of the manuscript. N.M. acknowledges the support provided through CREST, JST.

\appendix

\section{Preprocessing of the fMRI data\label{app:preprocessing}}

We used resting-state fMRI data publicly shared in the Human Connectome Project,  release WU-Minn S1200 \cite{Vanessen2012Neuroimage}.
The data were collected using a 3T MRI (Skyra, Siemens) with an echo planar imaging (EPI) sequence (TR, 0.72 s; TE, 33.1 ms; 72 slices; 2.0 mm isotopic; field of view, $208 \times 180$ mm) and T1-weighted sequence (TR, 2.4 s; TE, 2.14 ms; 0.7 mm isotopic; field of view, $224 \times 224$ mm). The EPI images were recorded in four runs
($\approx$ 15 min per run) while participants were instructed to relax while looking at a fixed cross mark on a dark screen. Each run yielded $1,200$ volumes (i.e., discrete time points).
We used such EPI and T1 images recorded from two adult participants 
arbitrarily selected from the 10 unrelated subject data set in the release
(one male of 26--30 years old and one male of 31--35 years old).
%
% 105115: M 31-35
% 118932: M 26-30

We preprocessed the EPI data obtained from each run in essentially the same manner as the conventional methods that we previously used for resting-state fMRI data \cite{Watanabe2013NatComm,Ezaki2017PhilTransRSocA} with SPM12 (www.fil.ucl.ac.uk/spm). After discarding the first ten images in each run, which yielded a time series of volumes of length $1,190$, we conducted realignment, slice timing correction, normalisation to the standard template (ICBM 152) and spatial smoothing (full-width at half maximum $=8$ mm). Afterwards, we removed the effects of head motion, white
matter signals and cerebrospinal fluid signals by a general linear model. Finally, we performed temporal band-pass filtering ($0.01$--$0.1$ Hz) and obtained resting-state whole-brain data.
We then extracted a time series of fMRI signals from each region of interest (ROI). The ROIs were defined as 4 mm spheres around their centre whose $N=264$ coordinates were determined in a previous study \cite{Power2011Neuron}. The
signals at each ROI were those averaged within the sphere.

Within each run and at each ROI, we subtracted the mean from the time series of fMRI signals. Then, we concatenated the fMRI data across the four runs to obtain a time series of length $L = 4,760$ at each ROI. We calculated the Pearson correlation coefficient between each pair of ROI to determine the correlation matrix for each participant.

\section{Hirschberger-Qu-Steuer algorithm\label{app:HQS}}

Given the covariance matrix, $\Sigma^{\rm org}$, the H-Q-S algorithm generates random covariance matrices, $\Sigma^{\rm HQS}$, satisfying the following conditions \cite{Hirschberger2007EurJOperRes}.
First, each on-diagonal element of a generated covariance matrix has the expected value that is equal to the average of the on-diagonal elements of the original covariance matrix.
Second, each off-diagonal element of a generated matrix has the expected value and the variance that are equal to the average and variance of the off-diagonal elements of the original matrix, respectively.
We did not implement a variant that also constraints the variance of the on-diagonal elements of a generated covariance matrix \cite{Hirschberger2007EurJOperRes} or
a fine-tuned heuristic variant of the algorithm \cite{Zalesky2012Neuroimage}.

Denote by $\mu_{\rm on}$ the average of the $N$ diagonal elements of the original covariance matrix. Denote by $\mu_{\rm off}$ and $\sigma_{\rm off}^2$ the average and variance of the off-diagonal elements of the original covariance matrix, respectively. We set
\begin{equation}
L^{\rm HQS} \equiv \max \left( 2, \lfloor \left(\mu_{\rm on}^2 - \mu_{\rm off}^2\right)/\sigma_{\rm off}^2\rfloor\right),
\end{equation}
where $\lfloor \cdot \rfloor$ is the largest integer that is smaller than or equal to the argument.
Then, we generate $N \times L^{\rm HQS}$ variables, denoted by $x_{i\ell}$ ($1\le i\le N$, $1\le \ell\le L^{\rm HQS}$), which independently obey the normal distribution with mean $\sqrt{\mu_{\rm off} / L^{\rm HQS}}$ and variance $-\mu_{\rm off} / L^{\rm HQS} +
\sqrt{\mu_{\rm off}^2 / (L^{\rm HQS})^2 + \sigma_{\rm off}^2 / L^{\rm HQS}}$. The H-Q-S algorithm sets
\begin{equation}
\Sigma^{\rm HQS}_{ij} = \sum_{\ell=1}^{L^{\rm HQS}} x_{i\ell} x_{j\ell}\quad
(1\le i, j\le N).
\end{equation}

The expectation of the samples generated by the H-Q-S algorithm is given by
$\langle\Sigma^{\rm HQS}\rangle_{ij} = \delta(i,j)\mu_{\rm on} + \left[1-\delta(i,j)\right]\mu_{\rm off}$ and $\langle\rho^{\rm HQS}\rangle_{ij} = \delta(i,j) + \left[1-\delta(i,j)\right]\mu_{\rm off}/\mu_{\rm on}$.

\section{Correlation matrices based on random matrix theory\label{app:MG}}

In this section, we explain null models $\langle \rho^{\rm MG2}\rangle$
and $\langle \rho^{\rm MG3}\rangle$ in Ref.~\cite{Macmahon2015PhysRevX}.

A given correlation matrix is decomposed as
\begin{equation}
\rho^{\rm org} = \sum_{i=1}^N \lambda_i \bm u_{(i)} \bm u_{(i)}^{\top},
\end{equation}
where $\lambda_i (\ge 0)$ is the $i$th largest eigenvalue and $\bm u_{(i)}$ is the corresponding normalised column eigenvector of $\rho^{\rm org}$.
Correlation matrix $\langle \rho^{\rm MG2}\rangle$
preserves the eigenmodes corresponding to small noisy eigenvalues
and is given by
\begin{equation}
\langle \rho^{\rm MG2}\rangle = \sum_{i=1; \lambda_i \le \lambda_+}^N \lambda_i \bm u_{(i)} \bm u_{(i)}^{\top},
\end{equation}
where
\begin{equation}
\lambda_+ = \left(1 + \sqrt{\frac{N}{L^{\rm MG}}}\right)^2
\end{equation}
and $L^{\rm MG}$ is the number of data points based on which the pairwise correlation is calculated.
Although $\langle \rho^{\rm MG2}\rangle$ is not a correlation matrix because its diagonal elements are not equal to 1, it does not affect the subsequent network analysis, which usually discards the diagonal elements \cite{Macmahon2015PhysRevX}.
Correlation matrix $\langle \rho^{\rm MG3}\rangle$ preserves the largest eigenmode in addition to the noisy eigenmodes and is given by
\begin{equation}
\langle \rho^{\rm MG3}\rangle = \lambda_1 \bm u_{(1)} \bm u_{(1)}^{\top} +
\sum_{i=2; \lambda_i \le \lambda_+}^N \lambda_i \bm u_{(i)} \bm u_{(i)}^{\top}.
\end{equation}

\section{White-noise model\label{app:white}}

We generated correlation matrices from independent white noise as follows.
For each node, we first generated a time series of length $L^{\rm WH}$, where each element obeys the standard normal distribution that is independent across time and nodes. Then, we calculate the Pearson correlation between the time series at node $i$ and that at node $j$ to define the $(i, j)$ element of the correlation matrix. As $L^{\rm WH}$ grows, the correlation matrix approaches the identity matrix owing to the law of large numbers. The H-Q-S model that happens to have $\mu_{\rm off} = 0$ is a special case of the white-noise model, where $L^{\rm WH} (= L^{\rm HQS})$ is typically small. As is the case for our configuration model, the value of $L^{\rm WH}$ affects the distribution of observables and hence the $P$ value when comparing a given correlation matrix and randomised correlation matrices. We set $L^{\rm WH}=N$.

\section{Definition of $C^{\rm wei,O}$\label{app:C^{wei,O}}}

The clustering coefficient for weighted networks proposed by Onnela and colleagues is given by \cite{Onnela2005PhysRevE}
\begin{equation}
C^{\rm wei,O} = \frac{1}{N} \sum_{i=1}^N C_i^{\rm wei,O}.
\label{eq:C^{wei,O}}
\end{equation}
In Eq.~\eqref{eq:C^{wei,O}}, the local clustering coefficient at node $i$, denoted by $C_i^{\rm wei,O}$, is given by
\begin{equation}
C_i^{\rm wei,O} = \frac{1}{k_i(k_i-1)} \sum_{\substack{1\le j,\ell\le N\\ j,\ell\neq i}} \frac{(w_{ij} w_{i\ell} w_{j\ell})^{1/3}}{\max_{i^{\prime}j^{\prime}} w_{i^{\prime}j^{\prime}}},
\label{eq:C_i^Onnela}
\end{equation}
where the edge weight $w_{ij} = \rho_{ij}$ if $\rho_{ij}$ is positive, and $w_{ij}=0$ otherwise. Factor $\max_{i^{\prime}j^{\prime}}w_{i^{\prime}j^{\prime}}$ normalises $C_i^{\rm wei,O}$ (and hence $C^{\rm wei,O}$) between zero and one and prevents it from scaling when $w_{ij}$ for all $1\le i, j\le N$ is multiplied by the same constant.

\section{Definition of $C^{\rm cor,M}$\label{app:C^{cor,M}}}

The partial mutual information is a nonlinear correlation measure given by \cite{Frenzel2007PhysRevLett}
\begin{equation}
I(\tilde{X}_j, \tilde{X}_{\ell} \mid \tilde{X}_i) = h(\tilde{X}_j, \tilde{X}_i) + h(\tilde{X}_{\ell}, \tilde{X}_i) - h(\tilde{X}_i) - h(\tilde{X}_j, \tilde{X}_{\ell}, \tilde{X}_i),
\label{eq:partial mutual info}
\end{equation}
where $\tilde{X}_i$, $\tilde{X}_j$ and $\tilde{X}_{\ell}$ are the random variables on nodes $i$, $j$ and $\ell$, respectively, and $h$ is the (joint) entropy. For example, $h(\tilde{X}_i) = \sum_{\tilde{x}} p(\tilde{x})\log_2 p(\tilde{x})$, where $p(\tilde{x})$ is the probability with which $\tilde{X}_i = \tilde{x}$, and $h(\tilde{X}_j, \tilde{X}_i) = \sum_{\tilde{x}, \tilde{x}^{\prime}} p(\tilde{x}, \tilde{x}^{\prime}) \log_2 p(\tilde{x}, \tilde{x}^{\prime})$, where $p(\tilde{x}, \tilde{x}^{\prime})$ is the probability with which $(\tilde{X}_j, \tilde{X}_i) = (\tilde{x}, \tilde{x}^{\prime})$. Under the assumption that the random variables on nodes $i$, $j$ and $\ell$ obey a multivariate Gaussian distribution, the entropy values in Eq.~\eqref{eq:partial mutual info} are simplified to \cite{Rieke1997-1999book,CoverThomas2006book,Frenzel2007PhysRevLett}
\begin{equation}
h(\tilde{X}_{\alpha_1}, \ldots, \tilde{X}_{\alpha_d}) = \frac{d}{2}(1+\ln 2\pi) + \frac{1}{2}\ln \det \Sigma^{\prime}.
\label{eq:joint h for Gaussian}
\end{equation}
In Eq.~\eqref{eq:joint h for Gaussian}, $d$ is the number of random variables and $\Sigma^{\prime} = (\Sigma^{\prime}_{ij})$ is the $d\times d$ covariance matrix derived from $\tilde{X}_{\alpha_1}$, $\ldots$, $\tilde{X}_{\alpha_d}$, i.e., $\Sigma^{\prime}_{ij} = \langle \tilde{X}_{\alpha_i} \tilde{X}_{\alpha_j}\rangle$, where we recall that $\langle \cdot \rangle$ represents the expectation. By substituting Eq.~\eqref{eq:joint h for Gaussian} in Eq.~\eqref{eq:partial mutual info} and feeding 
the correlation matrix as a covariance matrix to Eq.~\eqref{eq:joint h for Gaussian},
one obtains
\begin{align}
I(\tilde{X}_j, \tilde{X}_{\ell} \mid \tilde{X}_i) =& \frac{1}{2}\left[ 
\ln\left(1-\rho^2_{ij}\right) + \ln\left(1-\rho^2_{i\ell}\right) \right.\notag\\
-& \left. \ln\left(1-\rho^2_{ij}-\rho^2_{i\ell}-\rho^2_{j\ell} + 2\rho_{ij}\rho_{i\ell} \rho_{j\ell}\right)\right].
\label{eq:partial mutual info 2}
\end{align}
We define the local clustering coefficient at node $i$ as
\begin{equation}
C_i^{\rm cor,M} = \frac{\sum_{\substack{1\le j< \ell\le N_{\rm ROI}\\ j,\ell\neq i}} \left|\rho_{ij} \rho_{i\ell} \right| \times
I(\tilde{X}_j, \tilde{X}_{\ell} \mid \tilde{X}_i)}
{\frac{1+\ln 2\pi}{2} \sum_{\substack{1\le j< \ell\le N_{\rm ROI}\\ j,\ell\neq i}} \left|\rho_{ij} \rho_{i\ell}\right|}.
\label{eq:C_i^M}
\end{equation}
The denominator ensures $C_i^{\rm cor,M}$ to range between zero and one. The global clustering coefficient, denoted by $C^{\rm cor,M}$, is given by 
\begin{equation}
C^{\rm cor,M} = \frac{1}{N} \sum_{i=1}^N C^{\rm cor,M}_i.
\label{eq:C^M}
\end{equation}

\section{Parameter estimation by root finding\label{app:root finding}}

We present a procedure to calculate the precision matrix that maximises the entropy of $p(X)$ while respecting
\begin{equation}
\int \Sigma^{\rm con}_{ii}\; p(X) {\rm d}X = \Sigma^{\rm org}_{ii}
\label{eq:C_ii conservation}
\end{equation}
and
\begin{equation}
\int \sum_{j=1; j\neq i}^N \Sigma^{\rm con}_{ij}\; p(X) {\rm d}X = \sum_{j=1; j\neq i}^N \Sigma^{\rm org}_{ij},
\label{eq:strength conservation}
\end{equation}
where $1\le i\le N$.

Consider the precision matrix given by Eq.~\eqref{eq:precision matrix 2}.
Equations~\eqref{eq:C_ii conservation} and \eqref{eq:strength conservation} imply that
\begin{equation}
\Sigma \begin{pmatrix} 1\\ \vdots\\ 1\end{pmatrix} = 
\begin{pmatrix}
\sum_{j=1}^N \Sigma^{\rm org}_{1j}\\ \vdots\\ \sum_{j=1}^N \Sigma^{\rm org}_{Nj}
\end{pmatrix}
\equiv
\begin{pmatrix} u_1\\ \vdots \\ u_N \end{pmatrix}.
\end{equation}
Therefore, we obtain
\begin{equation}
\Sigma^{-1} 
\begin{pmatrix} u_1\\ \vdots \\ u_N \end{pmatrix}
= 
\begin{pmatrix} 1\\ \vdots\\ 1\end{pmatrix}.
\label{eq:constraint u}
\end{equation}
By combining Eqs.~\eqref{eq:precision matrix 2} and \eqref{eq:constraint u}, one obtains
\begin{equation}
f_i \equiv \alpha_i u_i + \beta_i \sum_{j=1}^N u_j + \sum_{j=1}^N \beta_j u_j - 1 = 0,
\label{eq:f_i=0}
\end{equation}
where $1\le i\le N$.

Equation~\eqref{eq:C_ii conservation} yields
\begin{equation}
\Sigma^{\rm org}_{ii} = \int \Sigma^{\rm con}_{ii}\; p(X) {\rm d}X = \frac{1}{L} \sum_{\ell=1}^N \int x_{i\ell}^2\; p(X) {\rm d}X = \Sigma_{ii} = \frac{{\rm Co}(i,i)}{\det \Sigma^{-1}},
\label{eq:C_ii conservation 2}
\end{equation}
where ${\rm Co}(i,i)$ is the $(i, i)$ cofactor of $\Sigma^{-1}$.
A straightforward calculation yields
\begin{equation}
\det \Sigma^{-1} =  \left(\prod_{\ell=1}^N \alpha_i\right)\times 
\left[ 1 + 2 \sum_{\ell=1}^N \frac{\beta_\ell}{\alpha_\ell} - \sum_{1\le \ell < \ell^{\prime}\le N} \frac{(\beta_\ell-\beta_{\ell^{\prime}})^2}{\alpha_\ell \alpha_{\ell^{\prime}}} \right].
\label{eq:det precision matrix}
\end{equation}
Therefore, the $(i, i)$ cofactor of $\Sigma^{-1}$ is given by
\begin{equation}
{\rm Co}(i,i) =  \left(\prod_{\substack{\ell=1\\ \ell\neq i}}^N \alpha_i\right)\times 
\left[ 1 + 2 \sum_{\substack{\ell=1\\ \ell\neq i}}^N \frac{\beta_\ell}{\alpha_\ell} - \sum_{\substack{1\le \ell < \ell^{\prime}\le N\\ \ell, \ell^{\prime}\neq i}} \frac{(\beta_\ell-\beta_{\ell^{\prime}})^2}{\alpha_\ell \alpha_{\ell^{\prime}}} \right].
\label{eq:Co precision matrix}
\end{equation}
By combining Eqs.~\eqref{eq:C_ii conservation 2}, \eqref{eq:det precision matrix} and \eqref{eq:Co precision matrix}, one obtains
\begin{align}
g_i \equiv& \alpha_i (1-\alpha_i v_i) \left[ 1 + 2 \sum_{\ell=1}^N \frac{\beta_\ell}{\alpha_\ell} - \sum_{1\le \ell < \ell^{\prime}\le N} \frac{(\beta_\ell-\beta_{\ell^{\prime}})^2}{\alpha_\ell \alpha_{\ell^{\prime}}} \right]
- 2\beta_i + \sum_{\ell=1}^N \frac{(\beta_i - \beta_\ell)^2}{\alpha_\ell}\notag\\
=& 0,
\label{eq:g_i=0}
\end{align}
where $v_i\equiv \Sigma^{\rm org}_{ii}$ and $1\le i\le N$.

Given $u_i$ and $v_i$ ($1\le i\le N$), functions $f_i$ and $g_i$ ($1\le i\le N$) define a system of $2N$ nonlinear equations for $2N$ unknowns, $\alpha_i$ and $\beta_i$ ($1\le i\le N$). We attempted to solve it using a MATLAB in-built function for root finding and an in-house implementation of the Newton-Raphson method. However, neither method could find the root, presumably because of the rugged landscape of $f_i$ and $g_i$ as a function of $\alpha_i$ and $\beta_i$. Rewriting Eqs.~\eqref{eq:f_i=0} and \eqref{eq:g_i=0} in terms of $\gamma_i\equiv 1/\alpha_i$ ($1\le i\le N$), which makes Eqs.~\eqref{eq:f_i=0} and \eqref{eq:g_i=0} polynomials in terms of $\beta_i$ and $\gamma_i$ ($1\le i\le N$), did not help.

%\bibliography{../citations}


\newpage
\clearpage

\newpage
\clearpage

\newpage
\clearpage

\newpage
\clearpage

\newpage
\clearpage

\newpage
\clearpage

\newpage
\clearpage

\newpage
\clearpage

\newpage
\clearpage

\begin{table}[htbp]
  \centering
\caption{Clustering coefficients. The clustering coefficient is denoted by $C$. For each type of randomised correlation matrices, the average and standard deviation based on $10^3$ samples are shown.}
\label{tab:C results} 
\begin{tabular}{lZYYcZYYY}
    \hline
    \hline
    \multicolumn{1}{c}{\multirow{2}[4]{*}{Null model\hspace{1em}}} & &\multicolumn{2}{c}{$C^{\rm wei,O}$} & &&  \multicolumn{2}{c}{$C^{\rm cor,M}$} \bigstrut\\ \cline{2-4} \cline{6-8} 
	&   $C$ & $Z$ & $P$ &\multirow{1}{1em}{\ } &  $C$ & $Z$ & $P$\bigstrut\\
    \hline
Motivation 	 & $0.284$ (empirical) 	 & 	 &&  	 & $0.031$ (empirical)\\
Configuration & $0.503 \pm 0.024$ & $-9.23$ & $<10^{-3}$  && $0.022 \pm 0.001$ & $6.20$ & $<10^{-3}$ \\ 
H-Q-S & $0.335 \pm 0.034$ & $-1.51$ & $0.130$     && $0.029 \pm 0.003$ & $0.59$ & $0.557$ \\
White-noise & $0.111 \pm 0.015$ & $11.26$ & $<10^{-3}$  && $0.001 \pm 0.000$ & $869.28$ & $<10^{-3}$ \\ \\
fMRI1 	 & $0.096$ (empirical) 	 & 	 && 	 & $0.013$ (empirical)\\
Configuration & $0.138 \pm 0.004$ & $-10.11$ & $<10^{-3}$  && $0.003 \pm 0.000$ & $85.90$ & $<10^{-3}$   \\
H-Q-S & $0.127 \pm 0.005$ & $-6.29$ & $<10^{-3}$   && $0.023 \pm 0.000$ & $-59.50$ & $<10^{-3}$  \\
White-noise & $0.078 \pm 0.005$ & $3.66$ & $<10^{-3}$    && $0.000 \pm 0.000$ & $23346.41$ & $<10^{-3}$\\ \\
fMRI2 	 & $0.101$ (empirical) 	 & 	 && 	 & $0.015$ (empirical)\\
Configuration & $0.147 \pm 0.004$ & $-11.11$ & $<10^{-3}$  && $0.003 \pm 0.000$ & $100.56$ & $<10^{-3}$  \\
H-Q-S & $0.119 \pm 0.005$ & $-3.34$ & $0.001$      && $0.017 \pm 0.000$ & $-19.99$ & $<10^{-3}$  \\
White-noise & $0.077 \pm 0.005$ & $4.54$ & $<10^{-3}$    && $0.000 \pm 0.000$ & $25741.87$ & $<10^{-3}$\\ \\
Japan 	 & $0.413$ (empirical) 	 & 	 && 	 & $0.027$ (empirical)\\
Configuration & $0.613 \pm 0.008$ & $-25.38$ & $<10^{-3}$ && $0.026 \pm 0.001$ & $1.32$ & $0.188$ \\ 
H-Q-S & $0.425 \pm 0.014$ & $-0.89$ & $0.376$     && $0.024 \pm 0.001$ & $3.77$ & $<10^{-3}$ \\
White-noise & $0.077 \pm 0.005$ & $62.77$ & $<10^{-3}$  && $0.000 \pm 0.000$ & $48473.21$ & $<10^{-3}$\\ \\
US 	 & $0.328$ (empirical) 	 & 	 && 	 & $0.024$ (empirical)\\
Configuration & $0.508 \pm 0.007$ & $-25.96$ & $<10^{-3}$ && $0.023 \pm 0.000$ & $2.80$ & $0.005$ \\
H-Q-S & $0.362 \pm 0.012$ & $-2.79$ & $0.005$     && $0.022 \pm 0.001$ & $4.82$ & $<10^{-3}$ \\
White-noise & $0.075 \pm 0.005$ & $51.94$ & $<10^{-3}$  && $0.000 \pm 0.000$ & $59738.94$ & $<10^{-3}$\\ 
    \hline
    \hline
    \end{tabular}
\end{table}

\newpage
\clearpage

\thispagestyle{empty}
\begin{table}[htbp]
  \centering
\caption{Community detection when the configuration null model, $\langle \rho^{\rm con}\rangle$, is given as input. For each null model, the mean and standard deviation of the modularity values based on $10^3$ samples of randomised networks are shown.}
\label{tab:comm config} 
\setlength{\tabcolsep}{3pt} 
\begin{tabular}{lYZYY}
    \hline
    \hline
    \multirow{2}[4]{*}{Null model\hspace{1em}} & \multicolumn{2}{c}{\centering $Q$}  & \multirow{2}{*}{$Z$} & \multirow{2}{*}{$P$} \\ \cline{2-3} 
	& Original & Random\\
    \hline
    Motivation\\
    $\langle \rho^{\rm HQS}\rangle$ & $0.15$ & $0.09 \pm 0.02$ & $2.72$ & $0.007$ \\
    $\langle \rho^{\rm MG1}\rangle$ & $0.88$ & $0.88 \pm 0.01$ & $0.85$ & $0.204$ \\
    $\langle \rho^{\rm MG2}\rangle$ & $0.98$ & $0.97 \pm 0.00$ & $3.18$  & $<10^{-3}$ \\
    $\langle \rho^{\rm MG3}\rangle$ & $0.12$ & $0.22 \pm 0.01$ & $-7.00$ & $1.000$ \\ \\
    fMRI1\\
    $\langle \rho^{\rm HQS}\rangle$ & $0.61$ & $0.17 \pm 0.01$ & $35.37$ & $<10^{-3}$\\
    $\langle \rho^{\rm MG1}\rangle$ & $1.10$ & $0.89 \pm 0.00$ & $54.62$ & $<10^{-3}$\\
    $\langle \rho^{\rm MG2}\rangle$ & $0.11$ & $1.02 \pm 0.00$ & $-966.06$ & $1.000$\\
    $\langle \rho^{\rm MG3}\rangle$ & $0.11$ & $0.50 \pm 0.01$ & $-36.92$ & $1.000$\\ \\
    fMRI2\\
    $\langle \rho^{\rm HQS}\rangle$ & $0.75$ & $0.20 \pm 0.01$ & $41.89$ & $<10^{-3}$ \\
    $\langle \rho^{\rm MG1}\rangle$ & $1.18$ & $0.87 \pm 0.01$ & $61.17$ & $<10^{-3}$ \\
    $\langle \rho^{\rm MG2}\rangle$ & $0.13$ & $1.04 \pm 0.00$ & $-452.25$ & $1.000$\\
    $\langle \rho^{\rm MG3}\rangle$ & $0.13$ & $0.56 \pm 0.01$ & $-36.99$ & $1.000$\\ \\
    Japan\\
    $\langle \rho^{\rm HQS}\rangle$ & $0.08$ & $0.04 \pm 0.01$ & $7.76$ & $<10^{-3}$\\
    $\langle \rho^{\rm MG1}\rangle$ & $0.99$ & $0.99 \pm 0.00$ & $-0.19$ & $0.582$\\
    $\langle \rho^{\rm MG2}\rangle$ & $0.01$ & $1.00 \pm 0.00$ & $-25328.61$ & $1.000$\\
    $\langle \rho^{\rm MG3}\rangle$ & $0.01$ & $0.09 \pm 0.00$ & $-17.07$ & $1.000$\\ \\
    US\\
    $\langle \rho^{\rm HQS}\rangle$ & $0.10$ & $0.05 \pm 0.01$ & $9.29$ & $<10^{-3}$\\
    $\langle \rho^{\rm MG1}\rangle$ & $0.99$ & $0.99 \pm 0.00$ & $-0.00$ & $0.482$\\
    $\langle \rho^{\rm MG2}\rangle$ & $0.01$ & $1.00 \pm 0.00$ & $-27404.80$ & $1.000$\\
    $\langle \rho^{\rm MG3}\rangle$ & $0.01$ & $0.11 \pm 0.00$ & $-19.89$ & $1.000$\\
    \hline
    \hline
    \end{tabular}
\end{table}

\begin{table}[htbp]
  \centering
\caption{Community detection for empirical correlation matrices. For each null model, the mean and standard deviation of the modularity values based on $10^3$ samples of randomised networks are shown.}
\label{tab:comm empirical} 
\begin{tabular}{lYZYY}
    \hline
    \hline
    \multirow{2}[4]{*}{Null model\hspace{1em}} & \multicolumn{2}{c}{\centering $Q$}  & \multirow{2}{*}{$Z$} & \multirow{2}{*}{$P$} \\ \cline{2-3} 
	& Original & Random\\
    \hline
    Motivation\\
    $\langle \rho^{\rm con}\rangle$ & $0.20$ & $0.03 \pm 0.02$ & $10.09$ & $<10^{-3}$ \\
    $\langle \rho^{\rm MG2}\rangle$ & $0.99$ & $0.98 \pm 0.00$ & $1.58$ & $0.044$  \\
    $\langle \rho^{\rm MG3}\rangle$ & $0.22$ & $0.30 \pm 0.02$ & $-3.38$ & $1.000$ \\ \\ 
    fMRI1\\                      
    $\langle \rho^{\rm con}\rangle$ & $0.95$ & $0.05 \pm 0.01$ & $83.31$ & $<10^{-3}$ \\
    $\langle \rho^{\rm MG2}\rangle$ & $1.82$ & $1.84 \pm 0.23$ & $-0.10$ & $0.493$ \\
    $\langle \rho^{\rm MG3}\rangle$ & $1.04$ & $1.71 \pm 0.22$ & $-3.01$ & $1.000$ \\ \\
    fMRI2\\                      
    $\langle \rho^{\rm con}\rangle$ & $1.36$ & $0.05 \pm 0.01$ & $110.67$ & $<10^{-3}$\\
    $\langle \rho^{\rm MG2}\rangle$ & $2.16$ & $1.70 \pm 0.18$ & $2.48$ & $0.013$  \\
    $\langle \rho^{\rm MG3}\rangle$ & $1.19$ & $1.62 \pm 0.19$ & $-2.26$ & $0.999$ \\ \\
    Japan\\                      
    $\langle \rho^{\rm con}\rangle$ & $0.03$ & $0.01 \pm 0.01$ & $3.95$ & $0.002$ \\
    $\langle \rho^{\rm MG2}\rangle$ & $1.00$ & $1.00 \pm 0.00$ & $4.69$ & $<10^{-3}$ \\
    $\langle \rho^{\rm MG3}\rangle$ & $0.04$ & $0.08 \pm 0.01$ & $-6.75$ & $1.000$ \\ \\
    US\\                         
    $\langle \rho^{\rm con}\rangle$ & $0.05$ & $0.01 \pm 0.01$ & $5.13$ & $<10^{-3}$ \\
    $\langle \rho^{\rm MG2}\rangle$ & $1.00$ & $1.00 \pm 0.00$ & $4.10$ & $<10^{-3}$ \\
    $\langle \rho^{\rm MG3}\rangle$ & $0.05$ & $0.11 \pm 0.01$ & $-7.91$ & $1.000$ \\
    \hline
    \hline
    \end{tabular}
\end{table}

\begin{table}[htbp]
  \centering
\caption{Consistency between the biological label of the nodes and the detected communities for the fMRI data.}
\setlength{\tabcolsep}{10pt} 
\begin{tabular}{lcYYYY}
    \hline
    \hline
    \multirow{2}[4]{*}{Null model\hspace{1em}} & \multirow{2}[4]{*}{Data} & \multirow{2}{*}{$P^{\rm emp}$} & \multirow{2}{*}{$P^{\rm rand}$} & $P^{\rm emp} - P^{\rm rand}$ & \multirow{2}{*}{$P^{\rm emp} / P^{\rm rand}$}\bigstrut \\ 
\hline
\multicolumn{3}{l}{All nodes} \\
\multirow{2}{*}{$\langle \rho^{\rm con}\rangle$}	& fMRI1 & $0.623$ & $0.316$ & $0.306$ & $1.968$ \\
						      	& fMRI2	& $0.673$ & $0.336$ & $0.337$ & $2.004$ \\
\multirow{2}{*}{$\langle \rho^{\rm MG2}\rangle$}	& fMRI1	& $0.725$ & $0.555$ & $0.170$ & $1.307$ \\
						      	& fMRI2 & $0.726$ & $0.520$ & $0.206$ & $1.397$ \\
\multirow{2}{*}{$\langle \rho^{\rm MG3}\rangle$}	& fMRI1 & $0.605$ & $0.434$ & $0.171$ & $1.393$ \\
                                                        & fMRI2 & $0.529$ & $0.362$ & $0.167$ & $1.460$ \\ \\
\multicolumn{3}{l}{Uncertain nodes removed} \\
\multirow{2}{*}{$\langle \rho^{\rm con}\rangle$}	& fMRI1 & $0.681$ & $0.316$ & $0.365$ & $2.155$ \\
						      	& fMRI2	& $0.723$ & $0.340$ & $0.382$ & $2.123$ \\
\multirow{2}{*}{$\langle \rho^{\rm MG2}\rangle$}	& fMRI1	& $0.778$ & $0.572$ & $0.206$ & $1.360$ \\
						      	& fMRI2 & $0.780$ & $0.551$ & $0.229$ & $1.415$ \\
\multirow{2}{*}{$\langle \rho^{\rm MG3}\rangle$}	& fMRI1 & $0.663$ & $0.467$ & $0.195$ & $1.418$ \\ 
						      	& fMRI2 & $0.570$ & $0.374$ & $0.196$ & $1.525$ \\
    \hline
    \hline
    \end{tabular}
    \label{tab:P^{emp}}
\end{table}
 
\end{document}